\documentclass[9pt,journal,compsoc]{IEEEtran}
\usepackage[normalem]{ulem}
\usepackage[usenames, dvipsnames]{color}
\usepackage{xcolor}
\usepackage{booktabs}

\definecolor{mypink}{HTML}{FFE7E6}
\definecolor{myyellow}{HTML}{FFF8D9}
\definecolor{mygreen}{HTML}{D9FFE7}
\definecolor{mypurple}{HTML}{E6E6FF}

\ifCLASSOPTIONcompsoc
  \usepackage[nocompress]{cite}
\else
  \usepackage{cite}
\fi
\ifCLASSINFOpdf
   \usepackage[pdftex]{graphicx}
\else
   \usepackage[dvips]{graphicx}
\fi
\usepackage{url}
\usepackage{hyperref}

\usepackage{enumitem}

\usepackage[usenames, dvipsnames]{color}

\newcommand{\etal}{et al.}
\newcommand{\inlineHeading}[1]{\vspace{0.04in}\noindent {\textbf{#1}}} 

\newcommand{\change}[1]{#1}

\newcommand{\task}[1]{T-{#1}}
\usepackage{xspace}
\newcommand{\toolname}[0]{{VIVA\xspace}}

\begin{document}
%
\title{\toolname{}: Virtual Healthcare Interactions Using Visual Analytics, With Controllability Through Configuration}
%
%
%
%

\author{J{\"u}rgen Bernard*, 
    Mara Solen*, 
    Helen Novak Lauscher, 
    Kurtis Stewart, 
    Kendall Ho,
    Tamara Munzner
    \IEEEcompsocitemizethanks{
        \IEEEcompsocthanksitem
        Bernard and Solen are joint first authors.
        \IEEEcompsocthanksitem
        J{\"u}rgen Bernard is with the University of Zurich, Department of Informatics and the Digital Society Initiative. E-mail: bernard@ifi.uzh.ch.
        \IEEEcompsocthanksitem
        Mara Solen and Tamara Munzner are with the University of British Columbia, Department of Computer Science. E-mail: marasolen@gmail.com, tmm@cs.ubc.ca.
        \IEEEcompsocthanksitem
        Helen Novak Lauscher, Kurtis Stewart, and Kendall Ho are with the University of British Columbia, Department of Emergency Medicine. E-mail: helen.nl@ubc.ca, kurtis.stewart@ubc.ca, kendall.ho@ubc.ca.}
    \thanks{Manuscript received XXX; revised XXX.}
}

%
%

\markboth{IEEE TRANSACTIONS ON VISUALIZATION AND COMPUTER GRAPHICS, ~Vol.~xx, No.~x, xxxx~xxxx}%
{Bernard \MakeLowercase{\textit{et al.}}: Virtual Healthcare Interactions Using Visual Analytics}
%



\IEEEtitleabstractindextext{%
  \begin{abstract}
       At the beginning of the COVID-19 pandemic, HealthLink BC (HLBC) rapidly integrated physicians into the triage process of their virtual healthcare service to improve patient outcomes and satisfaction with this service and preserve health care system capacity. We present the design and implementation of a visual analytics tool, VIVA (Virtual healthcare Interactions using Visual Analytics), to support HLBC in analysing various forms of usage data from the service. We abstract HLBC's \change{data and} data analysis tasks, which we use to inform our design of VIVA. \change{We also present the interactive workflow abstraction of Scan, Act, Adapt.} We validate VIVA's design through three case studies with stakeholder domain experts. We also propose the Controllability Through Configuration model to conduct and analyze design studies, and discuss architectural evolution of VIVA through that lens. It articulates configuration, both that specified by a developer or technical power user and that constructed automatically through log data from previous interactive sessions, as a bridge between the rigidity of \change{hardwired} programming and the time-consuming implementation of full end-user interactivity.  
    \textbf{Availability:} Supplemental materials at \change{\url{https://osf.io/wv38n}}.
  \end{abstract}

}

\maketitle

\IEEEdisplaynontitleabstractindextext

%



\IEEEraisesectionheading{\section{Introduction}}

\IEEEPARstart{I}{n} virtual healthcare, patients call clinicians to get health care advice and guidance by phone or video conference. This service can be efficiently provided by a healthcare system at relatively low cost, compared to in-person visits\change{, and} provides easy access for patients in any geographic location with no need for travel. Healthcare systems can use \change{it}
to triage for problems and guide patients towards next steps such as \change{seeking} further medical care immediately, or \change{seeking} care later, or \change{staying} home.

HealthLink BC (HLBC) is a longstanding virtual healthcare program that provides non-emergency advice from nurses across the entire province of British Columbia. The triage goal is friendly guidance in steering people to the right resource at the right time. At the start of the COVID-19 pandemic, there was an urgent need to avoid overwhelming emergency rooms; system costs for virtual visits are over an order of magnitude less than for in-person visits, 
and virtual visits minimize the patient hazard of exposure to contagious disease. The HealthLink BC Emergency iDoctor-in-assistance (HEiDi) initiative rapidly integrated physicians into the triage process to immediately provide more tailored advice through diagnostic judgement, spearheaded by our collaborators in the Digital Emergency Medicine (DigEM) group. Although such initiatives would normally take years of planning, it was rolled out at extreme speed in 6 weeks due to the COVID-19 crisis. Its goals were to provide high-quality patient care and satisfaction, and to preserve health care system capacity for delivering care. 

Usage data for HEiDi was collected and analyzed by health system experts at HLBC and DigEM to assess the program's effectiveness and efficiency, and to monitor strategic and operational concerns for the health system overall, both longstanding challenges and those specific to the COVID-19 crisis. Example analysis questions include: 
\textit{Which system usage patterns change when physicians \change{join} the triage system alongside nurses? Are patients satisfied with their experience in accessing health care through this initiative? Are there any shifts in the medical conditions that callers ask about, over time, or across the province?} 
The assessment criteria for the program were developed in parallel with project deployment due to the extreme speed of the rollout.
Early analysis was handled through a highly manual creation process, with monthly PowerPoint reports showcasing static charts hand created in Excel. 

Our collaborators wanted to explore the potential of a flexible visual data \change{discovery and} analysis workflow for this usage data, with the ability to drill down into details and also pull back to understand the big picture. 
The potential benefits they envisioned were to expedite analysis and decision-making by health system experts, and to enable better communication with many other stakeholders including clinicians, patients, and policymakers. 
Together with our collaborators, we designed and developed \toolname{}, short for \textbf{Vi}rtual healthcare \textbf{I}nteractions Using \textbf{V}isual \textbf{A}nalytics.

The contributions of this design study are three-fold. 
\begin{itemize}
\item We identify data, task\change{, and workflow} abstractions for the usage and results of virtual healthcare guidance delivered by clinicians within the public health domain. \change{We propose a \textbf{Scan, Act, Adapt} interactive workflow that provides a middle ground of flexibility, midway between general-purpose applications like Tableau or Excel and specialized bespoke tools arising from typical design studies.} 
\item We present the design, architecture, and implementation of
\toolname{}, a system for visual data analysis to expedite analysis and decision-making and enable communication by healthcare system experts. We validate \toolname{} through three case studies with domain experts. 
\item We propose a generalizable approach for conducting and analyzing design studies: the \textbf{Controllability Through Configuration} model, where configuration \change{bridges} and accelerate\change{s} the transition between hardwired programming by developers and a fully interactive end-user interface. 
\end{itemize}






\section{Process and Abstractions}
\label{sec:processAndAbstractions}

We now outline our research process \change{and goals}, and present abstractions for data, tasks\change{, and an interactive workflow}. 
\subsection{Process}
\label{sec:processAndAbstractions:process}

We largely followed the design study methodology (DSM) of Sedlmair \etal~\cite{SedlmairMM12}, featuring iterative refinement and close contact with domain experts. 
This design study entailed a collaboration between a design team of visual analytics researchers and the Digital Emergency Medicine (DigEM) research group at UBC, who in turn collaborate with HealthLink BC (HLBC). 
The HEiDi initiative took place after substantial previous engagement across these two organizations, so shared context and well-established communication pathways already existed. 
We followed a staged model~\cite{McLachlanMKN2008} to progressively gain access to both data from and target users at HLBC, in light of the well-documented challenges of achieving such access in highly regulated domains such as healthcare~\cite{CrisanGM2016}.

Casting these collaborators in roles,
the DigEM group leader (KH) was connector~\cite{SedlmairMM12}, translator~\cite{SedlmairMM12}, gatekeeper~\cite{SedlmairMM12}, and promoter~\cite{OppermannM2020}. There were two front-line analysts~\cite{SedlmairMM12}: one non-technical data consumer~\cite{KerznerBHM2015} (HL) and one technical data producer~\cite{KerznerBHM2015} and consumer (KS). These primary DigEM collaborators are also co-authors on this paper, as is common with design studies.
At HLBC, we gained access to a half-dozen people across several roles: data producer~\cite{KerznerBHM2015} and steward~\cite{CrisanGM2016}, report consumers, and front-line analysts. 

The project took place in \change{four} 
phases, over 3.5 
years. In the one-month Precondition phase, we conducted the \textit{winnow} and \textit{cast} DSM stages. 
\looseness=-1
The four-month Abstraction phase was for \textit{discover} and \textit{design}, followed by the six-month Build phase for iterative refinement of \textit{discover} and \textit{design}, and \textit{implement}. Finally, a multi-year Reflect phase encompassed \textit{deploy}, \textit{reflect}, and \textit{write}. 


The process involved multiple rounds of \change{expert engagement}. The initial requirements assessment was conducted through \change{9 unstructured meetings (1 to 1.5 hours each)} \change{with the DigEM experts} during the Precondition and Abstraction phases \change{(Months 1-5), yielding information about their goals, concrete low-level analysis questions, and data. The design team iteratively developed the data characterization and the core abstract tasks (\emph{Inspect}, \emph{Partition}, and \emph{Stratify}) through reflective synthesis informed by expert feedback.} 
\change{Three more supporting abstract tasks were gradually conceptualized as the design team gained an even deeper understanding of user needs during the iterative development in the Build phase (Months 6-12): \emph{Scan} in Month 6, \emph{Customize} in Month 7, and \emph{Curate} in Month 9.}

\change{In the Build phase we} gathered formative and summative feedback \change{abstractions and designs} through weekly one-hour meetings with DigEM front-line analysts, giving chauffeured demos to elicit feedback on the iterative refinement.  
\change{We} also held \change{held eight one-hour meetings with KH (translator and gatekeeper), whose analysis questions are distilled as Case Study 1}. 
\looseness=-1

\change{We obtained feedback and access permissions from HLBC personnel through targeted meetings throughout the phases.} In Month 3, we presented \change{initial} abstractions and mockups as part of our request for data access; it was granted in Month 4, \change{marking the transition} from Abstraction to Build \change{in Month 5}. In Months 10 and Month 11, we presented \toolname{} via chauffeured demos to HLBC, to obtain summative feedback on its suitability.

Training and deployment took place \change{early} in the final phase. In Month 12, we held a usage/training session with a DigEM non-technical front-line analyst (HL), summarized in Case Study 2. In Month 13, we held two usage/training sessions with an HLBC front-line analyst (EC), summarized in Case Study 3. In Month 14, we held a usage/training session with the translator (KH) to encourage further non-chauffeured usage.

Following month 14, scheduling \change{complexities led to intermittent activity with reflection and writing, so}
the final phase of the project was prolonged. \change{However, the additional time for reflection led to beneficial refinement in the framing of this publication, including the interactive \textit{Scan, Act, Adapt} workflow that we decided to present as an explicit abstraction, given its deviation from established information-seeking paradigms.
}

\begin{figure}[t]
  \centering
  \includegraphics[width=1.0\linewidth]{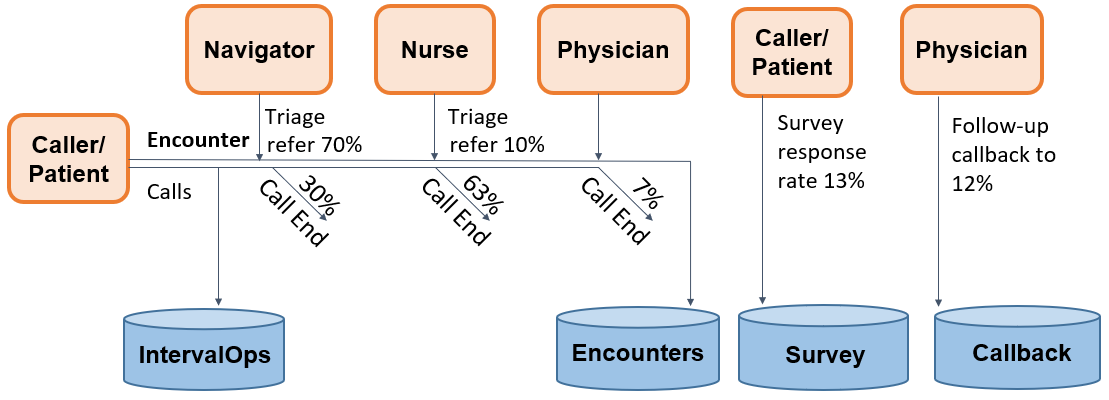}
  \caption{The HLBC patient call context with different actors (orange) along a patient encounter process, leading to data that is stored in four different data collections (blue). The Refer numbers are relative to the previous pipeline stage, and the Call End numbers are absolute for the whole collection.}
  \label{fig:patientCallContext}
\end{figure}

\subsection{\change{Goals}}
\label{sec:processAndAbstractions:goals}


The 
\change{DigEM}
experts quickly articulated seven high-level goals of the HEiDi program, which were well understood throughout their organizational structures. The first five are system-centric and the last two are patient-centric:

\begin{itemize}[noitemsep,topsep=1pt,parsep=1pt,partopsep=1pt]
    \item G1: Triage patients into appropriate care stream
    \item G2: Preserve capacity for health care system
    \item G3: Minimize utilization costs
    \item G4: Promote physician attachment for care continuity
    \item G5: Monitor HEiDi utilization and capacity
    \item G6: Ensure high-quality health outcomes
    \item G7: Ensure high-quality patient experience
\end{itemize}

The goal of preserving system capacity (G2) pertains specifically to the acute care services of emergency rooms and urgent care clinics, in the case of HEiDi by shifting system load to virtual care. Meanwhile in G4, \emph{attachment} refers to the long-term connection of patients to primary care (family) providers to provide continuity of care across non-acute settings. 

\subsection{Data Context and Characterization}
\label{sec:processAndAbstractions:dataContext}

Overall, \toolname{} includes and combines four tabular data sources, all of which emerge from the HLBC patient call context shown in Figure~\ref{fig:patientCallContext}.
An \emph{encounter} starts with a patient call.
A navigator forwards most of these calls to nurses\change{, who may refer some calls to the physicians, newly integrated in the process.}
\toolname{} includes only those encounter \change{data sent from nurses to physicians, representing} 7\% of total calls. Data that was recorded through the encounter process is stored the \emph{Encounters} data set with one item per encounter
\change{The \emph{IntervalOps} dataset captures aggregated statistics for HLBC call line activity at a 30-minute interval}.  
After the call, patients (callers) are immediately asked to answer an online survey (response rate of 13\%), with one item per respondent stored in the \emph{Survey} dataset. 
Two days later, physicians call back a subset (\~{}12\%) of the patients to \change{assess} well-being and treatment decisions, with one item per contacted patient stored in the \emph{CallBack} dataset. 
Figure~\ref{tab:dataCharacteristics} shows the total amount of data collected from April 2020 to March 2021, ranging from 27K encounters to 2K survey responses. It also shows the number of items per month added to each source; these datasets were collected and analyzed at regular intervals (typically 2-4 weeks). Any interpretation is limited to the data available during this timeframe and may not be consistent with current data or processes.

Figure~\ref{tab:dataCharacteristics} also shows the breakdown of attribute types across each source. The \emph{Encounters} dataset contains many categorical attributes including triage recommendations such as \emph{seek care now} or \emph{stay home} (which may change between the nurse and the physician consultation) and the type of health condition that led to the call such as \emph{Respiratory} or \emph{Gastrointestinal}, in addition to patient demographics such as age and gender, and the time/date of the call. The \emph{IntervalOps} aggregate statistics attributes are all quantitative, 
for example 
\change{total calls, average call duration, and answer rates per interval.}
The \emph{Survey} dataset includes many ordered attributes, such as self-rated patient satisfaction. \change{Though patients provided} free-form textual responses, \toolname{} does not support analyzing those. The \emph{Callback} dataset includes some follow-up information, such as whether patients \change{adhered to} triage recommendation, for a small sample of callers. 
Unlike the other three datasets, it is written down by the physicians and then manually transcribed by a DigEM staff member, so it is less clean than the others.
\change{All datasets contain temporal information, which \toolname{} uses} to present value changes over time.
Some time-oriented data could be used in its raw form (such as performance values or averages). 
\change{Others, like count-based data, require user-selected aggregation for temporal analysis.}

\change{Finally, Figure~\ref{tab:dataCharacteristics} shows the connections between \toolname{}'s four data sources and the analysis goals (Section~\ref{sec:processAndAbstractions:goals}), which were straightforward to identify.}







\begin{figure} 
\footnotesize
\renewcommand{\arraystretch}{1.2} 
\begin{tabular}{l|r|r|r|r|r|l}
 Source   &  Tot & Incr & C & O & Q & Goals\\
 \hline
Encounters   & 27K & 4000 & 22 & 0& 3 & G1-4, G6-7\\
IntervalOps & 23K & 3000 & 0 & 0 & 18 & G5\\
Survey & 2K & 200 & 12 & 8 & 4 & G2, G4, G6-7\\
CallBack & 4K & 500 & 7 & 1 & 2 & G4, G6-7\\
\end{tabular}
\vspace{2mm}
\caption{Data sources. Each source characterized by total size (for Apr 2020 through Mar 2021), monthly increase, attribute types (categorical, ordered, quantitative), and high-level analysis goals. }
\label{tab:dataCharacteristics}
\end{figure}



\subsection{\change{Task Abstraction}}


\change{Our interviews with the DigEM experts and analysis of their artifacts, including the early reports they manually generated, elicited a long list of low-level analysis questions.
We considered the connections between these concrete low-level data analysis questions and the data available from the four sources.} 
\change{Some of their} questions could not be answered without additional datasets that were not currently authorized for access by this group, for example whether eventual health outcomes were aligned with or divergent from the triage advice given in the call. However, \change{many} other questions were within the scope of the available data\change{, and were  thus within the scope of our task abstraction.}

\subsubsection{Core Tasks}
\label{sec:processAndAbstractions:taskAbstraction}

We
\change{identified} three core abstract tasks that would
serve to answer these questions: \textbf{Inspect}, \textbf{Partition}, and \textbf{Stratify}. 
To answer some basic questions, it is enough to \textbf{Inspect} a single attribute summarized across an entire time period, to check \change{the value} distribution across categorical or ordered attribute levels. 
\change{For example, inspecting the distribution of \texttt{gender} attribute levels (\texttt{male}, \texttt{female}), or patient satisfaction ratings (\texttt{from 1-5})}.
To answer other questions, it is important to \textbf{Partition} the attribute to show the temporal progression of how a single attribute changes over time. 
\change{For example, understanding daily counts of \texttt{yellow}-triaged patients over three months or call volumes in half-hour bins over weeks.}
Finally, multi-attribute analysis questions can be answered with the capability to \textbf{Stratify} across two or three attributes, such as both the gender and the age of patients.  

\change{\subsubsection{Supporting Tasks}}
\label{sec:processAndAbstractions:taskAbstraction:additionalTasks}

\change{Three additional abstract tasks} 
support core tasks; that is, they serve as the means for those ends. 
\change{We found that the analysts need to \textbf{Scan} through many attributes quickly,  \textbf{Customize} attributes for specific data analysis questions, and \textbf{Curate} attributes into groups of manageable size.} 

\change{These supporting tasks diverge
from common abstractions used in visualizations, and challenge several initial assumptions. Notably, we do not identify the need \change{for showing} geographic locations explicitly, for traditional data wrangling through pre-processing, for deriving new complex attributes, for a permanent categorization of attributes, or for global overviews. 
}

\inlineHeading{\change{Scan, Rather than Overview.}} \change{First, we specifically note that analysts needed to \emph{Scan}} through many attributes quickly \change{to simply} locate \change{an already known attribute of} interest \change{or check} for potential surprises in attribute distributions. However, the analysts did not need a global overview of all attributes simultaneously. \change{Through identifying several reasons for this unusual situation, we arrive at an interesting data/task abstraction that differs from many other visualization contexts. In addition to basic attribute types for tabular data (Sec.~\ref{sec:processAndAbstractions:dataContext}), our abstraction includes:} 
\begin{itemize}
    \item{\change{Attribute Familiarity}}. \change{Analysts} are deeply familiar with the \change{attributes long before using} \toolname{}. 
    \change{Traditional global-overview solutions would orient unfamiliar users to help them understand dataset structure, assess distributions, spot high-variance attributes, or check correlations.}    
    \change{However, our analysts do not need such answers; instead, they would question} whether their expert intuitions and expectations align with reality in a fine-grained way. 
    \item{Attribute vs.~Item Focus}. Overviews are most commonly used to show all items in a dataset. However, the three core tasks all focus on attributes and their distributions. None of them require diving into individual items. 
    \item{Few Attributes At Once}. The vast majority of \change{analysts' questions require only one or two, sometimes three attributes to answer them}. We originally posited that analysts would need complex multi-attribute queries, for example comparing female patients without primary physician attachment and with respiratory problems triaged to stay home between one month and the next. 
    \change{However, interviews and discussions revealed no such need.}
    \item{Sizeable Attribute Counts}. The union of the datasets contains nearly 90 original attributes, so it is not possible to show \change{detailed within-attribute distributions all at once. }
    In contrast, many bespoke global overviews are custom-designed for a moderate and fixed number of attributes.  
\end{itemize}

\inlineHeading{\change{On-the-Fly Customization of Attributes.}}
Eventually, we realized the importance of \change{enabling} analysts to \textbf{Customize} 
\change{an attribute dynamically rather than permanently modify or wrangle data or derive complex new attributes.}
\change{
\begin{itemize}[noitemsep,topsep=1pt,parsep=1pt,partopsep=1pt]
    \item{Routinely Re-Cleaned Data.} Rather than relying on upstream preprocessing and permanently modified data, analysts preferred adjusting data at the start of each session, refining attributes dynamically.
    \item{Specialized Needs Over General Wrangling}. Instead of altering attribute values directly, analysts manipulated categorical attributes by renaming levels, ignoring irrelevant ones, or merging them into a coarser granularity.
    \item{No Need for Advanced Derivations}. Analysts did not need sophisticated tools for deriving new attributes through mathematical formulas or complex combinations. Instead, they relied on simple operations like duplicating and renaming attributes.
\end{itemize}
}

\inlineHeading{\change{On-the-Fly Curation of Attribute Groups.}}
Another support task that we gradually understood is the need 
to \textbf{Curate} groups of attributes into more manageable chunks. 
\change{
\begin{itemize}[noitemsep,topsep=1pt,parsep=1pt,partopsep=1pt]
    \item{No Predefined Subsets.} Rather than predefined categorization, analysts assembled smaller, relevant subsets of original and derived attributes for each session.
    \item{No Disjoint Subsets}. A few highly-used attributes appeared across multiple subsets, making rigid partitioning ineffective for analysts. Instead, they tailored attribute groupings dynamically based on their immediate needs.
\end{itemize}
}



\inlineHeading{\change{No Need to See Geography.}}
Finally, geospatial analysis was not a core or even a supporting task for  healthcare system experts, in contrast to many other public health applications -- even though their data sources did include geographic region as part of patient demographics. 
\change{
\begin{itemize}[noitemsep,topsep=1pt,parsep=1pt,partopsep=1pt]
    \item{Regions Treated as Categorical}. Analysts only considered a few coarse-grained regions and were already familiar with major sectors of the province. There was no need to consult a geographic map.
    \item{Fine-Grained Geographic Data Unused}. While datasets contained detailed regional data, analysts never used it for low-level questions. 
\end{itemize}
}


\change{To summarize, analysts do not require global overviews, traditional pre-processing, complex derived attributes, permanent attribute categorization, or geographic visualizations.
Our data and task abstractions account for a sizeable number of attributes in total, all of them familiar, with only a few used at any given time. Moreover, there is no need for inspecting individual items.}

\begin{figure*}[t]
  \centering
  \includegraphics[width=\linewidth]{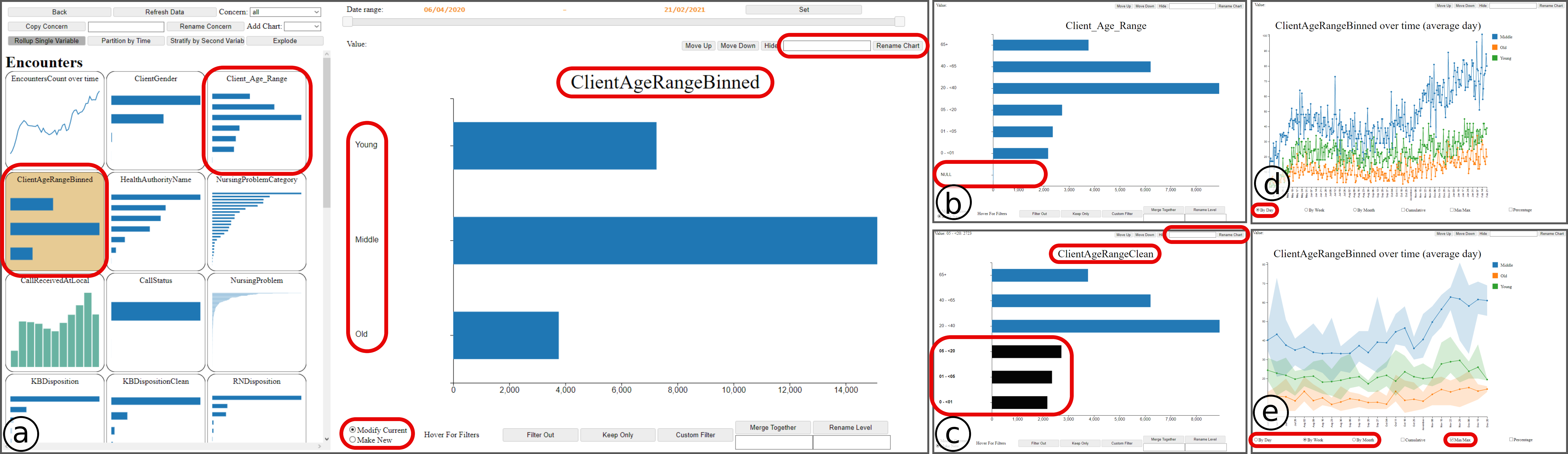}
  \caption{\toolname{} full interface, with Case Study 1 data. (a) The full interface of \toolname{} in Rollup mode after customizing an attribute: the new \texttt{ClientAgeRangeBinned} attribute shown in the Detail Panel has three levels (\texttt{Young}, \texttt{Middle}, \texttt{Old}) and its thumbnail is added to the current Concern attribute set, just after the original \texttt{Client\_Age\_Range} entry in the \change{Multiples} Panel grid. (b) Customization starts by filtering out the \texttt{NULL} level. (c) Customization continues with merging together the last three levels and Rename to \texttt{old}. (d) Partitioning over time triggers a switch from bar to line chart. (e) Changing to from day to week granularity aggregates data, as shown with min/max bars. }
  \label{fig:teaser}
\end{figure*}

\begin{figure*}[h]
  \centering
  \includegraphics[width=1.0\linewidth]{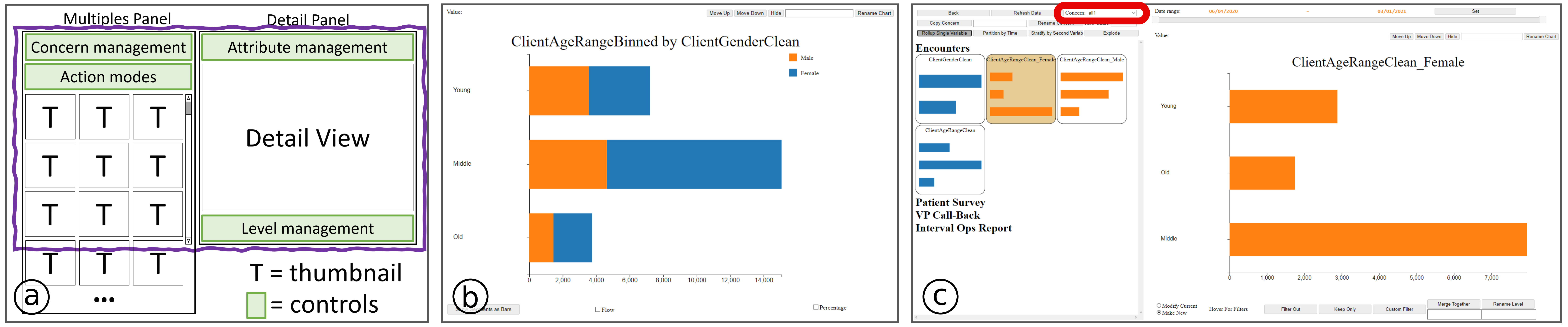}
  \caption{\toolname{} schematic, and Case Study 1 continuation. (a) Schematic interface diagram: the \change{Multiples} Panel on the left contains a scrollable grid of attribute thumbnails (T) and controls on top. The Details Panel on the right features a large detail view of the selected attribute(s), with controls above and below. The purple box depicts the visible screen area. (b) Stratifying age by gender yields a segmented bar chart. (c) Exploding age by gender reifies the stratification, creating a new Concern with one new derived attribute for each original bar, whose levels are the former segments.}
  \label{fig:Combo1}
\end{figure*}

\subsubsection{Task List}

In summary, the full list of core and supporting tasks is:
\begin{itemize} [noitemsep,topsep=1pt,parsep=1pt,partopsep=1pt]
    \item \textbf{\task{Inspect} \emph{Attribute}}:
    see single attribute's distribution of values
    \item \textbf{\task{Partition} \emph{Attribute}}: 
    analyze progression of attribute value distributions, partitioned across time period
    \item \textbf{\task{Stratify} \emph{Attributes}}: 
    compare the distribution of values between two or three attributes
    \item \textbf{\task{Scan} \emph{Attributes}}: 
    check many attributes quickly
    \item \textbf{\task{Customize} \emph{Attributes}}:
    manipulate levels of single attribute to make it more \change{useful}
    \item \textbf{\task{Curate} \emph{Attribute Group}}:
    identify and arrange meaningful subset of attributes
\end{itemize}

\subsection{Interactive Workflow Abstraction}
\label{sec:processAndAbstractions:designConsiderations}




\change{When reflecting on the design and use of \toolname{}, we further considered what aspects of our work could be a transferrable abstraction. In addition to the data and task abstractions that we present above, we realized that the workflow supported by our system is an interestingly different information-seeking paradigm than previously proposed ones, such as \emph{Overview First, Zoom and Filter, Details on Demand}~\cite{shneiderman96} or \emph{Search, Show Context, Expand on Demand}~\cite{vanham2009search}.}

\change{We propose the interactive workflow 
of \textbf{Scan, Act, Adapt}
as a generalization that could apply to other situations beyond our specific set of core and supporting tasks.  
In this workflow, analysts first identify a small set of attributes that are pertinent to a specific question, act upon that set to select relevant views, and often adapt those attributes on the fly to be more suitable for that investigation. In our case, \textit{Act} includes the three core tasks of \textit{Inspect}, \textit{Partition}, and \textit{Stratify}, while \textit{Adapt} encompasses both \textit{Customize} and \textit{Curate}. This workflow is an iterative analysis loop, where each round typically takes only a few minutes.} 
\change{Our workflow paradigm lies at an interesting region of the visualization system design space, at a middle ground of two different dimensions: flexibility and discovery.} 



{\inlineHeading{\change{Flexibility: Between Generalization and Specialization.}}
\change{\toolname{} is neither a fully general tool like Tableau nor as rigid as a fixed-purpose system. 
Instead, the \textit{Scan, Act, Adapt} workflow supports an intermediate level of focus, with iterative analysis loops 
that allow analysts to extract meaningful insights efficiently.} 
}
On one pole of the \change{flexibility dimension, highly general applications like Tableau or Excel allow analysts to answer a very broad variety of analysis questions, but require significant 
visualization expertise to use effectively because of that very power.
On the other pole, bespoke applications are designed and finely tuned for targeted task/data abstraction combinations, as commonly seen in design studies~\cite{SedlmairMM12}.}
Our design target is to lie 
in between the two. A system that could only answer the exact questions from these healthcare system experts about the exact datasets they have gathered would be 
\change{overly restrictive, limiting its broader applicability.}
\change{Instead, we identified a workflow abstraction} 
with sufficient power to encompass an interesting region of the analysis space, even as our goal is not to create a fully general application. 
\change{ 
}


\inlineHeading{\change{Discovery: Blending Exploration and Explanation.}} Although it is common to distinguish between exploration to discover unknowns and explanation to present the known as very high-level abstract goals~\cite{Munzner14}, these two actions are not mutually exclusive. 
We can consider them as opposite ends of a discovery spectrum, with room for hybrid possibilities in between. Many design studies~\cite{SedlmairMM12} target users who do exploratory data analysis on a daily \change{basis, possibly spending many hundreds} of hours on it. In contrast, healthcare system experts are not interested in open-ended exploration, due to the many other demands on their time. They do have the remit of generating understandable reports about their activities, on the presentation end of the spectrum. 
Although open-ended exploration is not a goal, they are aware that investing a few hours a month towards data analysis may provide them with an opportunity to present more compelling evidence of impact and value. \change{Our workflow abstraction thus targets the middle ground in this dimension as well.}

\section{Visual Interface}
\label{sec:VisualInterface}



\label{sec:VisualInterface:walkthrough}


The main page of \toolname{}, shown in Figure~\ref{fig:teaser}a, \change{features a} \emph{\change{Multiples} Panel} on the left, and the \emph{Detail Panel} on the right. The schematic diagram in Figure~\ref{fig:Combo1}a illustrates the overall structure. 

\subsection{\change{Scan: Multiples Panel}}
\label{sec:VisualInterface:multiples}

The \change{Multiples} Panel shows a scrollable grid of attribute thumbnails, divided into one section for each data source, allowing users to scan through many attributes quickly (\task{Scan}). 
The \change{Multiples} Panel 
shows only around a dozen attributes at a time;
\change{it is designed to support the \textit{Scan} stage of the interactive workflow through scrolling, and is not intended as a global overview.}
Our approach of showing within-attribute distributions as thumbnails is similar in spirit to previous work such as Voyager~\cite{Wongsuphasawat2016}, which shows a gallery of univariate summaries. 

All of the attribute thumbnails in the \change{Multiples} Panel show the \textit{Rollup} displays \change{(see Sec.~\ref{sec:VisualInterface:act})} except for the first one within each dataset section, which shows the number of items in that dataset partitioned over time. Original categorical attributes are shown in blue, original quantitative attributes in green, and derived attributes created with Explode mode are colored orange. 

\change{Which attributes are visible and where they appear in the grid is controlled through the \textit{Concern} mechanism for curating attribute groups (\task{Curate}), discussed further in Sec.~\ref{sec:VisualInterface:customize}.}



\subsection{\change{Act: Detail Panel}}
\label{sec:VisualInterface:act}

Selecting attribute cells from the \change{Multiples} Panel to act on changes the display in the Detail Panel to show the selected attribute(s) in that large view. The thumbnail and detail views for an attribute both show the same chart, but the large view has fully labelled axes, whereas the thumbnails only show the chart marks plus the attribute name on top. The chart type is automatically determined according to the attribute data type and the action mode chosen by \change{users}. 

\change{The}
controls for the four \emph{Action modes}\change{, which support the \textit{Act} stage of the workflow}, \change{are in the Multiples Panel just above the scrollable thumbnails section}:
\begin{itemize}[noitemsep,topsep=1pt,parsep=1pt,partopsep=1pt]
    \item \textbf{Rollup} of single attribute (\task{Inspect} \& \task{Customize})
    \item \textbf{Partition} attribute over time (\task{Partition})
    \item \textbf{Stratify} across multiple attributes (\task{Stratify})
    \item \textbf{Explode} attribute (\task{Stratify} \& \task{Customize} \& \task{Curate})
\end{itemize}

In \emph{Rollup} mode, as shown in Figure~\ref{fig:teaser}a-c, a single attribute can be inspected in detail (\task{Inspect}). 
Categorical and ordered attributes are shown with horizontal bar charts, so that the text labels for attribute levels are easy to read. These charts summarize the time-series information by showing the cumulative counts of instances for each level of the attribute across the entire active time range. The temporal range bi-slider control at the top of the Detail Panel supports temporal filtering for all modes. Any change to the time range is persistent and is immediately applied globally to all charts. This mode also supports customizing an attribute, as discussed in Sec.~\ref{sec:VisualInterface:customize}. 

Quantitative attributes are shown as histograms, with vertical orientation both so that they are easy to visually distinguish from the horizontal orientation of bar charts and because their numerical labels require less horizontal space for legibility. 

In \emph{Partition} mode, as shown in Figure~\ref{fig:teaser}d-e, a single categorical attribute is unrolled to show its temporal progression (\task{Partition}) using a multi-line chart with one curve for each attribute level. There are three levels of temporal granularity selectable in the level management panel, with day as the default as shown in Figure~\ref{fig:teaser}d. For the week-level and the month-level granularities the data is aggregated, and the summarized range can be shown with min-max bands as shown in Figure~\ref{fig:teaser}e. By default the absolute counts are shown, but \change{users} can also choose relative percentages. Similarly, the default is to show traditional time-series representations, but \change{users} can choose cumulative totals instead.

\begin{figure*}[t]
  \centering

  
  \includegraphics[width=1.0\linewidth]{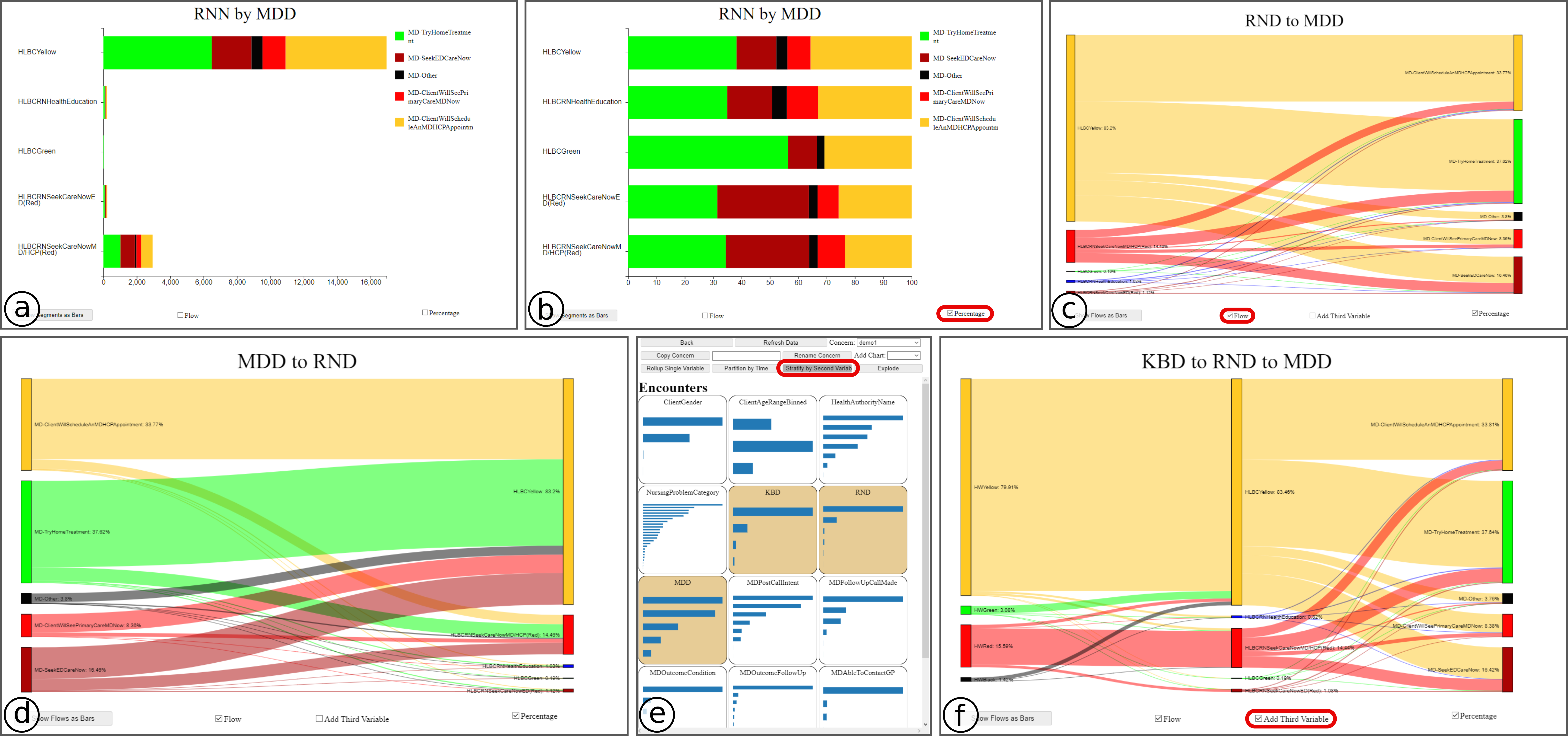}
  \caption{Case Study 1, Step 4: Stratification with segmented bars vs. Sankey flow diagrams, showing triage dispositions. (a) The initial visual encoding in Stratify mode is segmented bars, with bars for the first attribute selected (nurse triage, \texttt{RNN}) and segments for the second one (physician triage, \texttt{MDD}). (b) Selecting relative percentages reveals structure that is not visible with the default absolute counts. (c) If \change{analysts select} Flow, the chart type switches to a Sankey diagram, emphasizing the chronological order of the encounter when the patient is handed off from nurse to physician. (d) Reversing the order of selection instead emphasizes where the later recommendation differs from the previous one. (e) Flow mode allows \change{analysts to select} three attributes in the \change{Multiples} Panel for simultaneous display. (f) Seeing three attributes side by side in the Detail Panel highlights the progression of clinical judgements, from the generic recommendations (\texttt{KBD}) that are sometimes frequently overridden by nurses (\texttt{RND}), and in turn those are frequently changed by physicians (\texttt{MDD}).} 
  \label{fig:Combo2}
\end{figure*}

In \emph{Stratify} mode, as shown in Figure~\ref{fig:Combo1}b and Figure~\ref{fig:Combo2}, the Detail Panel shows information about two or three attributes simultaneously to support between-attribute analysis (\task{Stratify}). The order of attribute selection in the \change{Multiples} Panel determines the mapping of attribute values to visual encoding. The default chart type in this mode is a horizontal segmented bar chart. The attribute selected first determines the bars with one bar for each level, just as with the \emph{Rollup} mode. The levels of the second selected attribute dictate the segments nested within each bar. Relative percentages are an alternative to absolute numbers, as in Figure~\ref{fig:Combo2}b.

The user can instead choose to see a Sankey diagram, to focus on the flow between attributes, as shown in Figure~\ref{fig:Combo2}c. It is particularly appropriate for attributes linked through the temporal relationships illustrated in  Figure~\ref{fig:patientCallContext}, such as the progression of triage recommendations when an encounter progresses from nurse (\texttt{RNN}) to physician (\texttt{MDD}) consultation across different stages of the encounter process. The bars of the first attribute become the categories of the left axis, and the segments of the second attribute are the categories of the right axis. Switching to the opposite non-chronological order can also provide insight, for example by focusing on which physician decisions overrode the previous nurse decisions as shown in Figure~\ref{fig:Combo2}d. Users can add a third attribute with the flow display, as in Figure~\ref{fig:Combo2}e and Figure~\ref{fig:Combo2}f. The axis segments can also be sized by relative percentage in this variant. 

The \emph{Explode} mode, as shown in Figure~\ref{fig:Combo1}c and Figure~\ref{fig:UC3DispositionOverTime}, a stratification between two attributes is reified into a form that can subsequently be further inspected and manipulated. A new derived attribute is created for each bar (the levels of the first selected attribute), with levels corresponding to the segments (the levels of the second selected attribute). A new Concern is created containing only the two original attributes and the new derived ones. This mode addresses three tasks simultaneously: stratifying (\task{Stratify}), customizing (\task{Customize}), and curating (\task{Curate}) attributes. 

\subsection{\change{Adapt:} Customizing and Curating Attributes}
\label{sec:VisualInterface:customize}

\toolname{} supports \change{the \textit{Adapt} stage of the workflow through} customization within attributes and curation of attribute groups, allowing users to focus on which attributes and attribute levels are relevant for current analysis questions.


In \emph{Rollup} mode, the attribute being inspected can also be customized (\task{Customize}) by filtering levels out or keeping only certain levels, by merging levels together, and by renaming levels. Analysts first select which levels to act upon by clicking on bars or their labels in the Display Panel, then choose which actions to take with the controls at the top and bottom of the panel. Several steps of customization are shown in Fig~\ref{fig:teaser}a-c. 

When customizing attributes, the result of the within-attribute processing can either be to modify the attribute in-place (\emph{Modify Current}) or to create a new derived attribute (\emph{Make New}). When an attribute is first customized the mode is automatically set to \emph{Make New}, which derives a new attribute, whereas for subsequent operations it is reset to \emph{Modify Current}, which modifies the attribute in-place. The user can manually override these defaults through radio button controls at the bottom of the Display window. When a new derived attribute is created, its thumbnail is automatically inserted right after the original attribute's thumbnail in the current Concern grid.  

The \toolname{} interface supports attribute curation (\task{Curate}) through a general mechanism for grouping attributes into \textbf{Concerns}, which control what attributes are visible and where exactly they appear. 
Concerns can be managed through interaction with the interface. The grid of attribute thumbnails in the \change{Multiples} Panel shows the currently active Concern, which can be changed through a dropdown widget in the attribute management panel. 
Concerns created during an analysis session are automatically stored and loaded in subsequent ones.
The default Concern contains all attributes. Ten more specialized Concerns, determined \change{in consultation with} the domain experts, are also provided in the initial \toolname{} configuration. The full set of default Concerns and their constituent attributes are included in supplemental materials. 

\change{Sometimes} 
attribute curation is automatic, for example when a new attribute arises from customization or a new Concern is the result of an Explode action. It is also possible to manually curate any attribute when it is shown in the Detail Panel: it can be renamed (Figure~\ref{fig:teaser}c), explicitly moved around within the thumbnail grid, or removed from the current Concern using controls at the top of the panel. 

\change{The \textit{Adapt} workflow stage actively} encourages \change{analysts} to derive new customized attributes, so the number of attributes to show will gradually increase during an analysis session. \change{The Concern curation mechanism handles this increase.} 
\change{Since the Multiples panel is intended  to be scrolled while scanning rather than to present a global overview of all attributes side by side, this increase does not require any additional design changes to handle.}

\section{Case Studies}
\label{sec:caseStudiesAndResults}

We present three case studies 
\change{to illustrate \toolname{}'s utility}.


\subsection{Case Study 1: Chauffeured Demo (KH)}
\label{sec:caseStudiesAndResults:caseStudy1}

\change{\noindent{\textbf{Background:}}} This case study features \change{a selection of} analysis questions identified by KH, the leader of the DigEM research group who often served in the translator role~\cite{SedlmairMM12}, during our weekly meetings in the Build phase. 
\change{As KH attended these meetings only occasionally, he was not as familiar with the software as those who attended the meetings every week, however he did learn to use the tool during the Build phase.}
We distilled his core questions into a concise and compact chauffeured demo for a session in Month 10 to elicit summative feedback about \toolname{} suitability from HLBC stakeholders, with the visualization developer running the software itself in tandem with KH talking through his analysis.

\change{\noindent{\textbf{Age over time:}} KH began the demo by selecting relevant attributes from the \change{Multiples} Panel on the left (\task{Scan}) so that they appear in the Detail panel on the right (\task{Inspect}). He selects \texttt{Client\_Age\_Range}, which is renamed to \texttt{ClientAgeRangeBinned} and cleaned by merging levels together to reduce the granularity as shown in Figure~\ref{fig:teaser}a-c.}
\change{KH then disaggregates \texttt{ClientAgeRangeBinned} over time and views the data aggregated \emph{By Week} with \emph{Min/Max bands} to see trends over time as shown in Figure~\ref{fig:teaser}d-e. He is able to use the generated chart to point out} that middle-aged patients are the most frequent cohort across the entire time range, and increase substantially towards the end, although the young and old bins stay roughly constant. \change{He is then also able to narrow} the time range with the top bi-slider (Figure~\ref{fig:teaser}e) to check the increase interval more closely.

\change{\noindent{\textbf{Gender ratios:}} He cleans another attribute, naming it \texttt{ClientGenderClean}, then he} chooses to \emph{Stratify} the age attribute by the gender attribute (\task{Stratify}). \change{The generated chart, shown in Figure~\ref{fig:Combo1}b, makes the stand-out pattern clear; the proportion of female callers of middle age is high, contrasting with the near-equal gender split for young patients. The \emph{Explode} mode enables KH to generate new age attributes, including one filtered to only include female patients, as shown in Figure~\ref{fig:Combo1}c, to better show the distribution of the data.}



\begin{figure*}[t]
 \centering
  \includegraphics[width=1.0\linewidth]{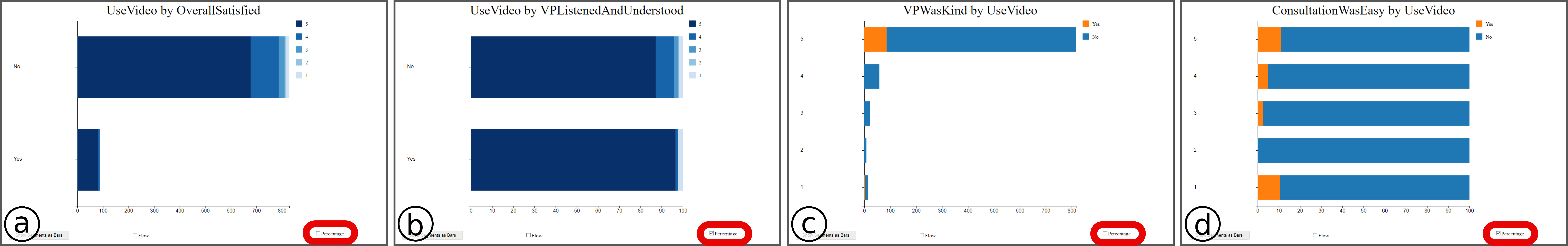}
  \caption{Case Study 2, Step 1. HN analyzes the satisfaction of callers when using video vs.~phone communication. To answer concrete data-centric questions, she uses the stratify mode (\task{Stratify}) to combine the video usage attributes with different user satisfaction scores, and checks both absolute numbers and relative percentages in each case. (a) If they used video, were patients more likely to be satisfied (absolute)? (b) If they used video, were patients more likely to feel the physician listened (percentages)? (c) Were the patients more likely to think the physician was kind if they used video (absolute)? (d) Were callers more likely to feel the consultation was easy if they used video (percentages)?}
  \label{fig:caseStudy2:T7}
\end{figure*}

\begin{figure*}[h]
  \centering
  \includegraphics[width=1.0\linewidth]{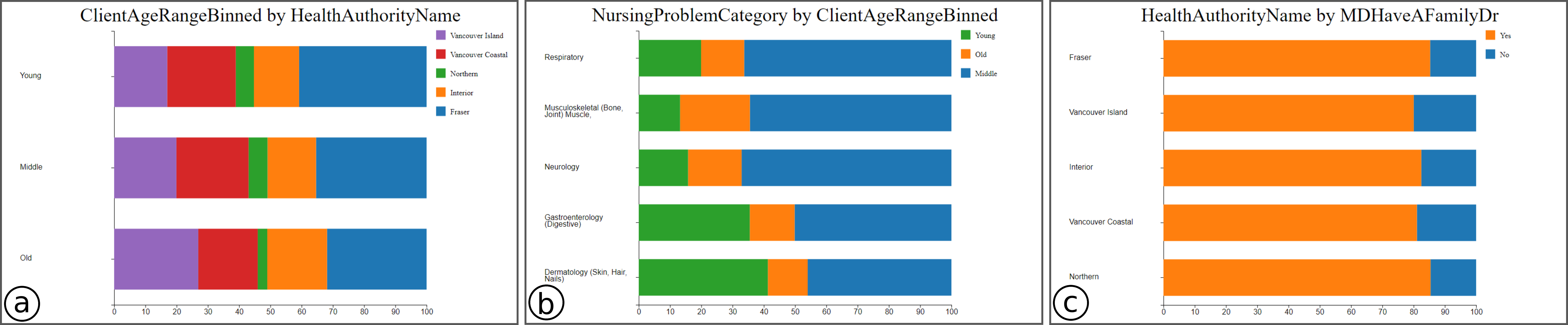}
  \caption{Case Study 2, Step 2. HN compares two attributes (\task{Stratify}). (a) On Vancouver Island (purple) the number of old callers is comparatively high, whereas in Fraser Valley (blue) the callers are relatively young. (b) By looking at the top five nursing problem categories, HN also identifies differences across young (green), middle (blue), and older (orange) callers. (c) In contrast, no considerable differences can be identified for different health authorities of having a primary doctor (orange) or not (blue).}
  \label{fig:caseStudy2:T7c}
\end{figure*}

\change{\noindent{\textbf{Triage flows:}}} 
KH moves on to analyze the attributes surrounding the most central question for the HLBC decision makers: the differential triage of patients (dispositions) performed by nurses (\texttt{RN}) vs.~physicians (\texttt{MD}). 
His goal is to gain insight into interprofessional team functioning and the prospect of mutual knowledge exchange for more harmonization of patient classification between the two professions in HLBC.
\change{He first \emph{stratifies}} nurse disposition by physician disposition, as shown in Figure~\ref{fig:Combo2}a, \change{and notices two bars are significantly bigger than the other three, making it much more difficult to see their distribution of segments. He switches} to relative distributions with \emph{Percentage}, as shown in Figure~\ref{fig:Combo2}b, \change{and sees that the distributions between all nursing distribution categories are roughly similar.}

KH switches to the \emph{Flow} mode to show the relationship between these two attributes differently, through a Sankey diagram (Figure~\ref{fig:Combo2}c-d). With the switch from segmented bars to flows, the HLBC stakeholders instantly identified the high number of state transitions from yellow (seek care) to green (stay home), meaning that in many cases the physician triage successfully allowed patients to avoid unnecessary in-person interactions. The HLBC stakeholders found this representation easier to reason about than the stacked bars, aligning with the reactions of the DigEM stakeholders in previous meetings. The flows in these Sankey diagrams nicely align with the chronological order of an HLBC call, where patients always speak to a nurse before a \change{physician.}

Finally, KH returns to the \change{Multiples} Panel, as shown in Figure~\ref{fig:Combo2}e, to add a third attribute to the \emph{stratification}: \texttt{KBD}, the guidance that would arise from simply following a handbook, which is recorded in the Encounter dataset before human triage judgements come into play. \change{Resulting from this addition is Figure~\ref{fig:Combo2}f, which leads to} a nuanced conversation between KH and the HLBC stakeholders about the meaning of this chart \change{due to}
intrinsically missing data: 
if the nurse triages to green (stay home), the call is not passed to a physician, so these nurse dispositions are not included in this dataset.


\change{\noindent{\textbf{Outcome:}} One important outcome showing the success of this chauffeured demo was the HLBC leadership team's commitment of a front-line analyst's time to learn and use this software}, which we present below as Case Study 3.


\subsection{Case Study 2: First Solo Use (HN)}
\label{sec:caseStudiesAndResults:caseStudy2}

\change{\noindent{\textbf{Background:}}} The second case study provides details about a \change{one hour} test drive from HN, a non-technical front-line analyst responsible for evaluating the efficacy of the HEiDi program. It is HN's first solo use of \toolname{}, \change{however she is familiar with the tool from} many previous rounds of chauffeured feedback demos in weekly meetings, \change{which she regularly attended}. Since the goal of the session was to understand the utility of the tool, rather than its usability, how-to questions were immediately answered by the visualization designers.
\looseness=-1
In advance of the test drive session, HN had prepared a list of dozens of data-centric questions covering three out of four data sources: Survey, Encounters, and Callback\change{, shaping} the story line of Case Study 2.


\change{\noindent{\textbf{Patient satisfaction:}}} HN intends to analyze many of the ordered attributes covering satisfaction ratings from the Patient Survey dataset, focused on the communication method of phone vs.~video. \change{She selects attributes (\task{Scan}) to analyze and cleans them (\task{Customize}).
Using both standard bars and normalized \emph{Percentage} mode,} HN then compares between several different attributes measuring satisfaction through 5-point Likert scales and their communication modality (\task{Stratify}), as shown in Figure~\ref{fig:caseStudy2:T7}. \change{She draws} useful conclusions, including the observation that only a single call resulted in the visually salient orange segment on the lowest bar representing the number of calls with the lowest satisfaction score (\texttt{1}), shown in Figure~\ref{fig:caseStudy2:T7}d. \change{From these charts, HN is quickly able to ascertain} a positive effect of video usage for caller satisfaction, which indicates that providing the possibility to have videotelephony was the right decision in many cases.



\begin{figure*}[t]
  \centering
  \includegraphics[width=1.0\linewidth]{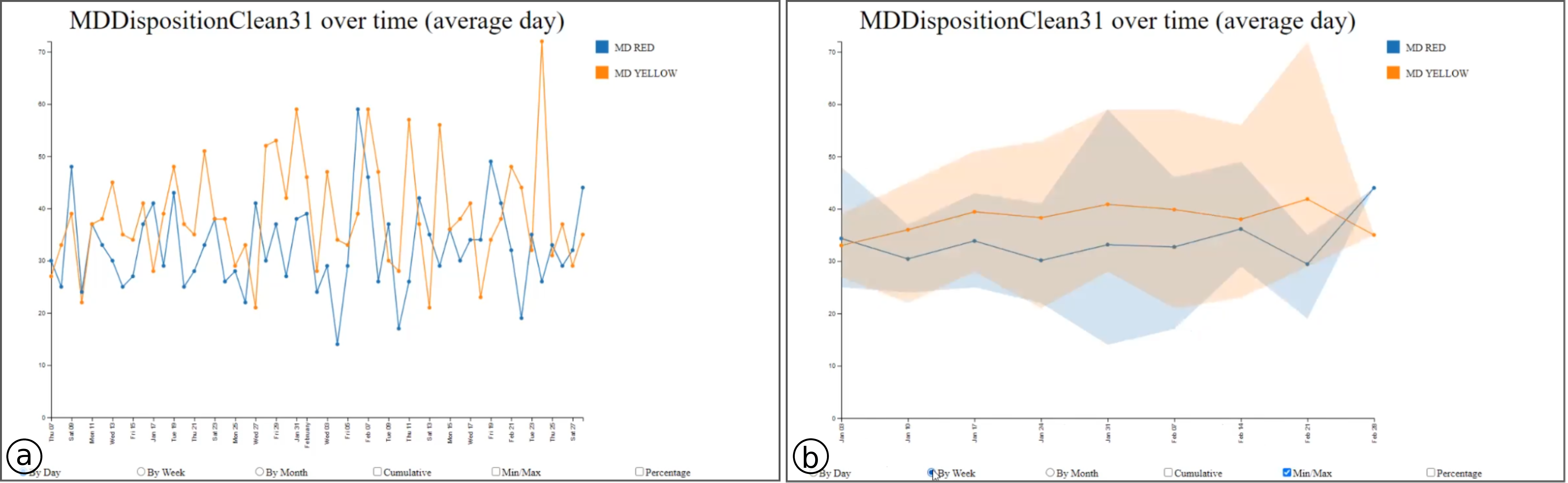}
  \caption{Case Study 3, Step 1: Two visualizations interactively created by EC to investigate the progression over time of encounters with recommendations to seek care soon or immediately, across the time interval of January and February 2021. The temporal progression does not show an increase of such calls, which is a good sign that current measures were preserving the capacity of the health care system. (a) By day. (b) By week, with min-max bands.}
  \label{fig:uc3T5MD}
\end{figure*}

\change{\noindent{\textbf{Aligning with existing knowledge:}}} HN then switches to the Encounter dataset \change{and makes some comparisons} using stratify mode (\task{Stratify}), \change{as shown in Figure~\ref{fig:caseStudy2:T7c}, to quickly verify that the patterns align with her existing knowledge.} She checks if the age of caller varies by health authority (Figure~\ref{fig:caseStudy2:T7c}a). She identifies that more elderly people are calling from Vancouver Island than in the Fraser region and Northern territory\change{;} she explicitly notes this pattern makes sense because it aligns with her knowledge of demographics in those regions. She then investigates the relationship between the top five nursing problem categories and patient ages (Figure~\ref{fig:caseStudy2:T7c}b). Young people suffer most from \texttt{Dermatology} and \texttt{Gastroenterology}, middle-aged callers report \texttt{Respiratory} and \texttt{Neurology} problems more often, and old people report \texttt{Musculoskeletal} disorders more often. Again, she finds these patterns align with her previous knowledge. Finally, she checks the proportion of patients attached to primary doctors across the five regional health authorities (Figure~\ref{fig:caseStudy2:T7c}c), and is pleased to see that there are no substantial differences.


\change{\noindent{\textbf{Outcome:}}} HN expresses satisfaction at what she could find using \toolname{} herself \change{as a non-technical analyst}, where she could investigate many aspects of the datasets very quickly and easily. \change{HN's session resulted in a very different analysis journey than the previous case study, showing the flexibility of the interactive \textit{Scan, Act, Adapt} workflow.}


\subsection{Case Study 3: New Technical Analyst (EC)}
\label{sec:caseStudiesAndResults:caseStudy3}

\change{\noindent{\textbf{Background:}}} In our final case study, we report on two \change{one-hour} sessions of use by EC, an HLBC technical front-line analyst who is very familiar with this data; he is the person who regularly sends this data to the DigEM researchers. His work entails analyzing data and generating reports for HLBC management. EC uses the combination of Excel templates and SQL queries to create static visuals that he incorporates into the PDF reports. Some management requests are for recurring reports showing weekly or monthly updates with the latest data. EC usually uses existing templates and scripts to quickly make the visual elements to incorporate into these routine reports, within 30 to 60 minutes. However, the workflow is slower if the data has changed format since the previous report was created. The other slow workflows arise from requests to create new kinds of visuals or do new kinds of analysis that are different from what he has produced before, which may require several days or even a week to complete. The time-consuming challenges for these more exploratory requests include retrieving new data, linking different SQL tables together, making sure that queries are right, checking the data, debugging formula problems, and generally ensuring that Excel has not produced unexpected results. EC is actively seeking better tools to support this process; to date, standard business intelligence tools have not been deemed to be a viable option because of many data management and privacy considerations. He considers interactivity to be a potential source of speedup to his analysis and report generation process, even though the final reports for management will be static PDF documents.

Before these sessions, EC had never previously interacted with the design team, \change{or seen a demo of \toolname{}.}
He had heard about \toolname{} from his HLBC colleagues and was enthusiastic to try it for himself. In the first session, he ran the software himself on our hardware through remote control of the screen sharing session. 
\looseness=-1
He then installed the software on his own computer, 
experimented further with it between sessions, and ran it on his own machine in the second session. In both cases, we observed and recorded his actions through screen sharing, while asking him to think aloud. We responded to any questions and made suggestions for what operations might be relevant for his expressed analysis interests, since our goal was to assess utility not usability. 

\change{\noindent{\textbf{Visual sanity check:}}} 
In the first session, EC's choices in how to use \toolname{} were very similar to the operations reported in the previous two case studies (cf. Figure~\ref{fig:teaser}): he conducted many sanity checks by scanning through attributes, visually inspected them, and frequently customized attributes to clean the data. He appreciated the benefits of being able to see everything all at once and clean the data on the fly with an interactive visual tool, \change{saying that was ``where it would save [him] a ton of time in doing these little breakdowns and combinations of fields and data''. He suggested} that it would be useful to be able to easily \change{export} the charts made with the tool, so we added a Print Chart button to save the chart \change{in the Detail Panel} as an SVG file before his second test drive. In this case study, we focus the second test drive session. 


\begin{figure*}[t]
  \includegraphics[width=1.0\linewidth]{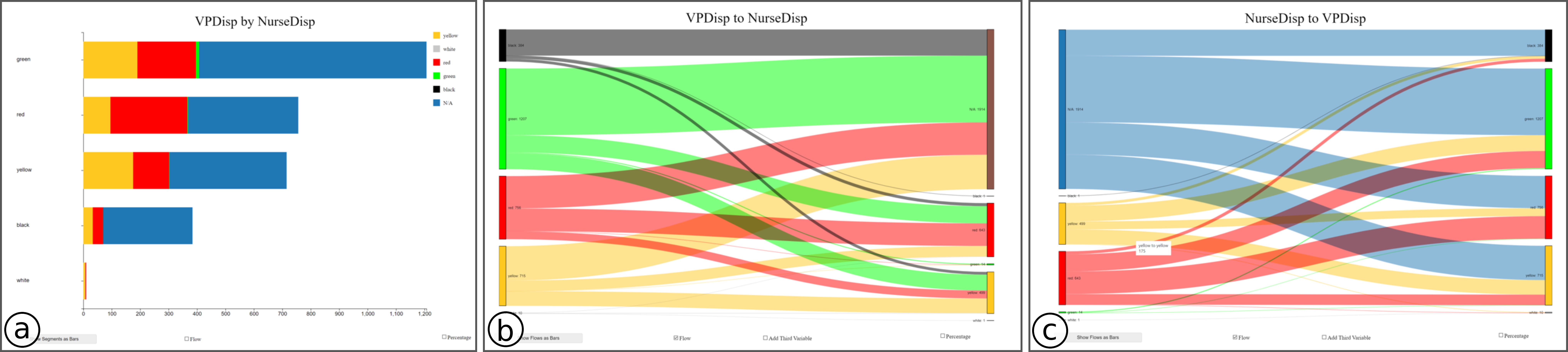}
  \caption{Case Study 3, Step 2: EC analyzes the disposition recommendations from physicians (VPs) versus nurses, using stratification. (a) The default segmented bar view emphasizes ratio discrepancies between the cases. (b) The flow view emphasizes the semantics that related decisions are made at different time points. (c) The flow view, with directionality reversed to match the flow of time within the encounter.}
  \label{fig:UC3VPcallbackNurseVPDisposition}
\end{figure*}

\begin{figure*}[h]
  \includegraphics[width=1.0\linewidth]{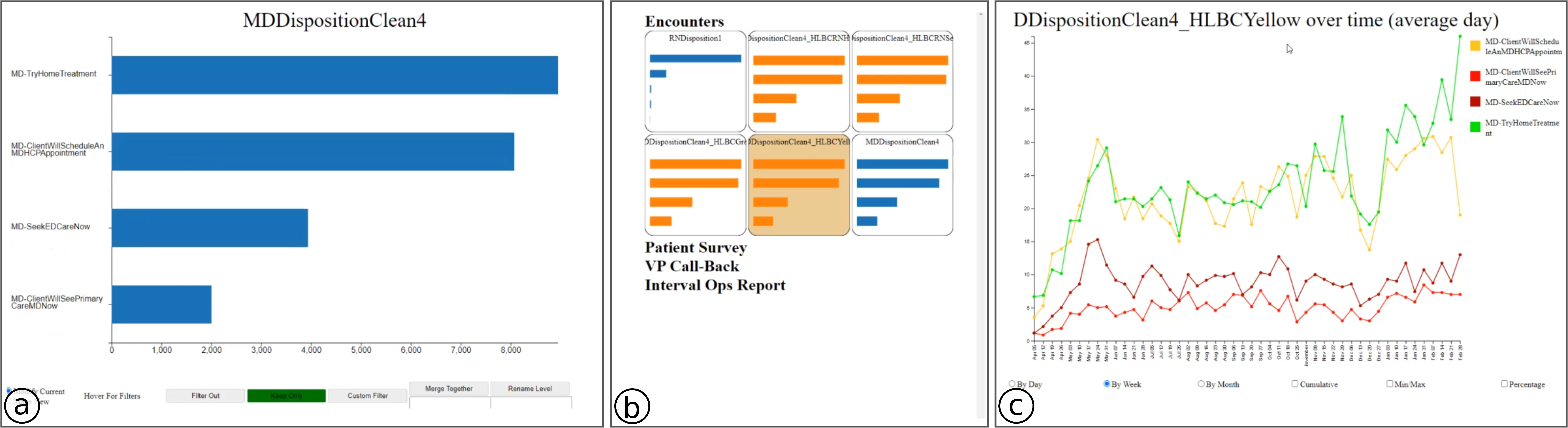}
  \caption{Case Study 3, Step 3: The Explode mode as applied in EC's workflow. (a) The cleaned \texttt{MDD} disposition attribute with four levels. (b) He explodes the \texttt{RNDisposition1} attribute by these four levels, leading to the four new attributes shown in orange. He then selects the new attribute where the nurse's disposition was to schedule an appointment. (c) He partitions the selected attribute's data over time.}
  \label{fig:UC3DispositionOverTime}
\end{figure*}

\change{\noindent{\textbf{Triage over time:}}} \change{He} recapitulates an analysis from the previous week that entailed a lengthy \change{and} slow workflow using his existing tools. In this analysis, he investigates the temporal patterns of encounter dispositions in the first two months of 2021, focusing \change{on} whether the dispositions that involve recommendations from the physicians that patients seek care soon or immediately are increasing over time or staying stable. He selects \change{and customizes the relevant attribute (\task{Customize}), filtering out the other dispositions} to keep only the three \change{relevant levels}. He then merges \change{and renames to arrive at the levels of \texttt{MD RED} and \texttt{MD YELLOW} which align with the HLBC conventions while maintaining simplicity.}


\change{\noindent{\textbf{Tracking disposition trends:}}}
He then partitions the derived attribute by time (\task{Partition}), and compares the \texttt{RED} and \texttt{YELLOW} disposition types as shown in Figure~\ref{fig:uc3T5MD}a, \change{before switching} to a weekly time granularity \change{with} min-max bands,
as in Figure~\ref{fig:uc3T5MD}b. He is pleased to see that no increasing trend of critical cases has happened in this time window, and clicks the Print Chart button to preserve the visual analysis result. EC expresses great enthusiasm for 
conducting all of these operations interactively and with visual support, noting that the speed \change{of \toolname{}'s interactive workflow} (under five minutes) saves him a lot of time compared to his previous workflow.  


\change{\noindent{\textbf{Callback dataset disposition:}}} Next, EC focuses on the Callback data \change{source.} He notices that the physician disposition attribute has eleven levels, which is surprisingly high given his knowledge of the data. EC \change{investigates} the attribute in detail (\task{Inspect}), and sees that some of these levels are differently encoded versions of the same semantics, such as \texttt{Red} or \texttt{High} meaning the same thing. He notes that this dataset was created by transcribing the handwritten notes of the physicians, and so needs extensive cleaning. He \change{cleans this attribute along with the nurse disposition attribute and stratifies them} together (\task{Stratify}) to understand which recommendations were overturned. \change{He} switches to the flow mode (Figure~\ref{fig:UC3VPcallbackNurseVPDisposition}b-c) \change{which he} finds easier to understand, expressing substantial enthusiasm for this visual representation (\textit{``I love this feature, by the way''}).


\change{\noindent{\textbf{Tracking disposition overrides:}}}
EC expresses interest in diving deeper into the cases where the nurse recommended that the patients schedule an appointment with their primary care doctor soon, and how the physicians choose to override or not override that recommendation. Based on this interest, the design team recommends that he use the Explode feature. He cleans two attributes from the Encounters data source, for physician and nurse dispositions. \change{The} physician disposition is shown in Figure~\ref{fig:UC3DispositionOverTime}a, with the long level names on the left associated with these dispositions. When he selects those two cleaned attributes \change{and} uses the Explode button, \toolname{} \change{shows the original two attributes in blue and the new exploded attributes in orange}, as shown in Figure~\ref{fig:UC3DispositionOverTime}b. \change{EC's domain expertise and familiarity with the datasets helps him in quickly understanding the names of the newly-generated attributes, but he is able to rename them as needed for reports or other communication.}

\change{\noindent{\textbf{Impact of physicians' expertise:}}}
EC now selects the exploded attribute that he is interested in: the physician dispositions in cases where the nurse suggested that the patient schedule an appointment. He partitions this new attribute over time to investigate changes to these decisions, as shown in Figure~\ref{fig:UC3DispositionOverTime}c. The multi-line chart that appears reveals \change{an interesting insight, as it indicates that} the additional expertise of the physicians who were added to HLBC's virtual healthcare service is both preserving capacity by suggesting that people with less critical conditions stay home and also potentially improving patient outcomes by suggesting that those whose conditions are deemed more critical get the care they need immediately. \change{He additionally notices increases in the green and yellow curves which contrast with the stability of the red curves, suggesting} that the number of encounters that result in a recommendation to seek urgent care is staying the same, despite an overall increase in encounters. Again, this result indicates that the physicians' expertise is successfully supporting the goals of HLBC.

\change{\noindent{\textbf{Outcome:}}} EC remarks that these insights were exactly what he was looking for, and concludes that using \toolname{} would save him a significant amount of time, up to multiple days per month, especially for conducting exploratory data analysis tasks \change{(``this analysis would definitely cut some time'')}.



\section{Architectural Evolution}
\label{sec:architecturalEvolution}


We first present the Controllability Through Configuration model, then analyze the evolution of \toolname{} through that lens, and finally discuss the software architecture of \toolname{}. 

\subsection{Controllability Through Configuration Model}
\label{sec:architecturalEvolution:controllabilitySpectrum}
\begin{figure}[t]
  \centering
  \includegraphics[width=1.0\linewidth]{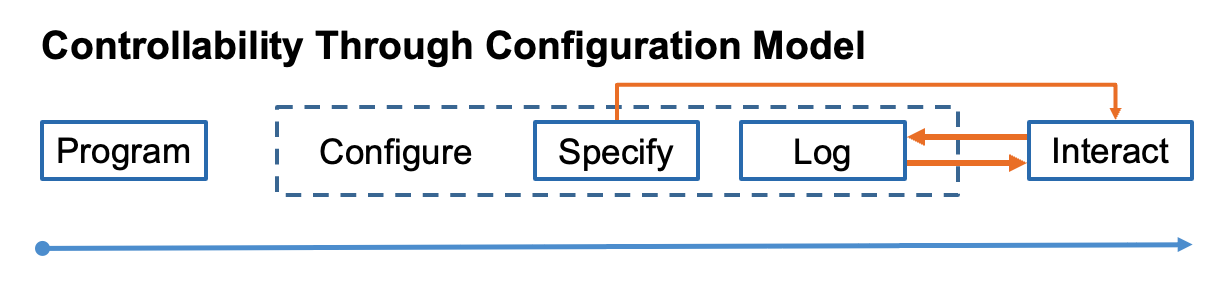}
  \caption{
  \change{Controllability through Configuration Model.}}
  \label{fig:ControlThroughConfig}
\end{figure}

Rapid development and feedback elicitation are common \change{goals} across many design studies, including our own: to effectively support HLBC as they rapidly implemented changes to their service, we needed to develop \toolname{} with equivalent speed. We identified a key strategy to accomplish these goals: the \change{balanced} use of \change{hardwired}, configured, and interactive components for system control. Through reflection, we built on this experience to develop generalizable approach for conducting and analyzing visualization design studies with speed and flexibility: the \textit{Controllability Through Configuration} model, shown in Figure~\ref{fig:ControlThroughConfig}. It features
multiple methods of control: programming, interactivity, and configuration. 

Programming is a powerful control mechanism with vast flexibility for the developer up through compile time, but hardwired code is rigid at runtime\change{, lacking adaptivity for users}. Iteration requires another round of coding from the developer, which takes time.  

Interactivity is very common in design studies, and is often added early as possible in iterative development cycles. Interactivity has obvious benefits, offering many degrees of freedom for runtime controllability. However, the design and development of interactive interfaces requires substantial coding effort for the developer, often requiring multiple iterations of usability tuning for a single feature update, and may even require users to spend time learning about how to use them. 

Configuration is an alternative method of control, 
without the need for recompilation. We define a \textit{configurable} system as one that responds to files at startup time. We argue that it is valuable to consider how configuration can serve as a bridge along a spectrum between programming and interactivity. 

With \textit{Specified} configuration, these files are explicitly created by a person. That person could be the developer, which allows a tight coupling between coding and configuration, especially at early stages of iterative development. This method allows for extremely fast turnaround because changes can be made on the fly during discussions. For example, we were able to conduct multiple rounds of iteration during one-hour weekly meetings between the developers and the stakeholders, rather than needing to wait for the next week's meeting to elicit feedback from the stakeholders on the suitability of an idea proposed the previous week. This approach combines some of the speed of paper prototyping, where feedback can be obtained on quick sketches within minutes, with some of the power of data sketches~\cite{LloydD2011}, where users \change{can} see concrete results on familiar datasets rather than being asked to simply reason about what might happen with their data without computational support. At a more mature phase of the development cycle, such files could also be created by power users. 

With \textit{Logged} configuration, the configuration files are the result of previous sessions with the software where user interaction was logged. The developer chooses what kinds of functionality should be logged so that user choices persist across sessions. For example, in \toolname{}, we logged all data customization so that modified and derived data attributes would be available for use in subsequent sessions. However, \change{we chose not} to log interactive browsing choices because our navigation model was quite straightforward; in systems with more complex navigation possibilities, logging that history could provide helpful support in maintaining user orientation. Of course, logged configuration can only occur after some level of interactive control provided to users, so there is a chronological dependency between these two controllability methods.  

\begin{figure*}[t]
  \centering
  \includegraphics[width=1.0\linewidth]{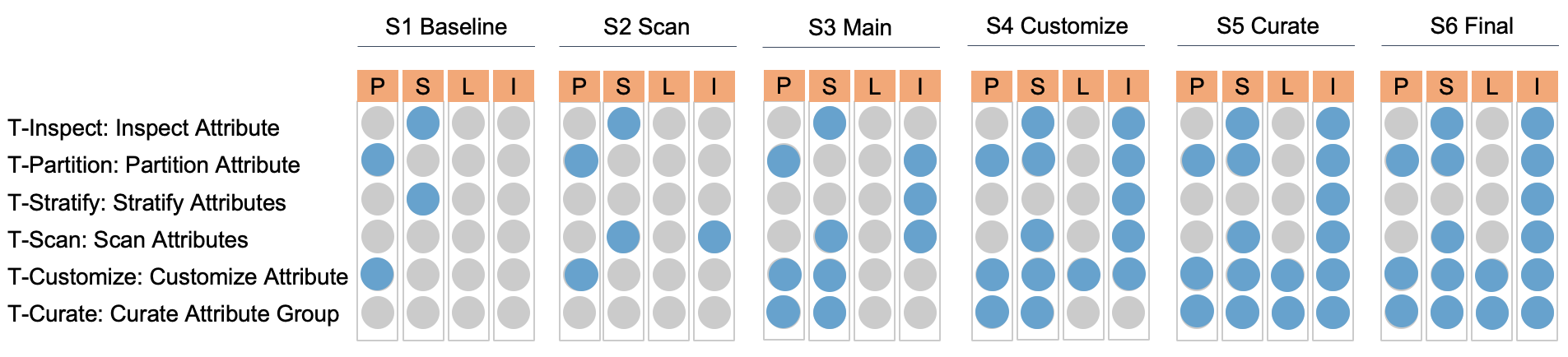}
  \caption{Controllability Evolution of \toolname{}: Six stages of the \toolname{} iterative design process showing which methods on the Controllability Spectrum supported which tasks: \change{(P)} Programming by the developer, \change{(S)} Specification through files created manually by the developer, \change{(L)} Logs of previous interactions by \change{users}, or \change{(I)} Interaction by \change{users} in the current session. By the final iteration, all tasks are supported interactively, but both specified and logged configuration was still used extensively.
  }
  \label{fig:ControlHistory}
\end{figure*}

We argue that the explicit characterization of controllability methods in the Controllability Through Configuration model provides an interesting lens through which to analyze existing design studies, providing descriptive power, as we show in the next section. We also provide this model in hopes that it will have generative power, to guide the development of future design studies.

\subsection{Controllability Evolution of \toolname{}}
\label{sec:architecturalEvolution:controllabilityEvolution}

This design study of \toolname{} is an example of the speed and the power of using both specified and logged configuration approaches to controllability in addition to the usual methods of developer programming and end-user interaction. At early stages of the iterative refinement process, we were able to quickly iterate through configuration rather than programming. At middle stages, we first added interactivity for only the most crucial aspects of functionality and used configuration as a bridging shortcut for other aspects, again to accelerate the iteration to mature versions where all tasks were supported through interactive controls. At later stages, we could easily make user choices persistent by adding targeted logging to generate configuration files automatically, since we already had a mechanism for importing them at startup. 

Figure~\ref{fig:ControlHistory} shows the evolution of \toolname{} according to the four controllability methods of programming, specified configuration files, logged configuration files, and interactivity at six milestone stages during its development. We now discuss each stage in turn; screenshots and all configuration files for each stage are included in Supplemental Section S1. 

\subsubsection{S1 Baseline}
\label{sec:architecturalEvolution:controllabilityEvolution:s1Baseline}
The initial stage of \toolname{} was developed very rapidly, within a few days of starting the Build phase, to sanity-check the data. The underlying \toolname{} code supported rendering several basic chart types: line charts, histograms, and both simple and stacked bar charts. The only user interaction supported was a slider to select the date range across which data was shown. 

This very early baseline instantiation already used configuration as a key design principle. A configuration file fully controlled the display of individual charts (\task{Inspect}) through one-line specifications for each chart that indicated which chart type to use, which dataset attributes to display within it, and what title to use for it in the interface. Stratifying one attribute by another (\task{Stratify}) was also handled through configuration by specifying two rather than one display attributes with the bar chart type, resulting in a stacked rather than simple bar chart. Partitioning over time (\task{Partition}) was mostly handled through programming: although specifying the chart type of \texttt{line} implicitly selected this action mode, the knowledge of which attribute within each dataset represented temporal data was hardwired. The configuration file syntax was ad hoc and complex, with considerable overloaded functionality, including the ability to compute and show aggregate statistics for time-series data as line charts. 

Seven charts were specified in the configuration file, chosen to replicate several charts from a non-interactive dashboard that the HEiDi team had created as a single PowerPoint slide, hand-assembled from charts that they manually created in Excel. It was manually generated by the developer. 
 

\subsubsection{S2 Scan}
\label{sec:architecturalEvolution:controllabilityEvolution:s2Scan}
The second stage of \toolname{} expanded its capabilities to show all dataset attributes (\task{Scan}). 
It featured the first use of interactively linked views with a drill-down design. In the \change{Multiples} View on the left were small multiples, in one large scrollable panel showing every attribute from every dataset with a small simple chart without labels or axes. When \change{analysts} interactively selected one attribute, it was shown in a large Detail View on the right (\task{Inspect}). 

The \change{Multiples} View was specified through a single configuration file specifying one chart per line, with straightforward and streamlined functionality: which dataset, which single attribute, and which attribute type. The set of attribute types was \texttt{categorical}, \texttt{numerical}, \texttt{time}, \texttt{datetime}, \texttt{percent}, \texttt{list}, and \texttt{freeform} text. The small number of ordered attributes, all for Likert scales, were initially mapped to categorical for simplicity. Numerical types additionally required a \texttt{units} field, such as \texttt{seconds} or \texttt{count}. \texttt{Numerical} and \texttt{percent} attributes were automatically displayed as vertical histograms, and all others shown as horizontal bar charts. 

The 75 attributes across the 4 datasets were enumerated in the configuration file, leaving out only 2 identifier attributes deemed to be absolutely irrelevant by the stakeholders. 
One more was quickly removed from the display by simply commenting it out, during a weekly meeting with stakeholders when they remarked that it was uninteresting to inspect. Neither stratification (\task{Stratify}) nor partitioning (\task{Partition}) were supported.


\subsubsection{S3 Main}
\label{sec:architecturalEvolution:controllabilityEvolution:s3PartStrat}
The third stage filled out the main functionality of \toolname{} with support for all six tasks. It provided interactive control of partitioning (\task{Partition}) and stratification (\task{Stratify}) through linked views, by supporting two selections within the \change{Multiples} View.  Stratification (\task{Stratify}) was fully handled through direct interaction: selecting two attributes showed a stacked bar chart in the right Detail View, with the selection order dictating which attribute was used for bars vs.~segments. If the second selection was a chart for a \texttt{time} attribute, the Detail View showed a line chart of the first attribute partitioned over time. 

This stage also introduced simple automatic data customization (\task{Customize}), through configuration to specify when multiple levels of a categorical or ordered attribute should be merged together, such as \texttt{Yes} and \texttt{yes} and \texttt{Y}. This customization was specified with a second manually generated configuration file containing one attribute per line. Browsing with the previous iterations of \toolname{} had immediately revealed many different kinds of noise in the datasets to both the stakeholders and the developers. This fast and low-stakes mechanism provided clear provenance of exactly what was merged, and we used it only for the situations where there was unequivocal agreement that the merge should happen in all cases. 
 
Curation (\task{Curate}) was also born at this stage, with support for interactively switching between different groups of attributes in the \change{Multiples} Panel. The intent was to help \change{analysts} quickly find attributes of interest by creating meaningful subsets of attributes relevant for specific analysis questions, rather than scrolling through the full set of 74 attributes visible in the \change{Multiples} Panel.
We identified ten Concerns, with under 20 attributes each, through stakeholder discussions. Examples include patient satisfaction, call volumes, or the patient journey through triage categories. We also added explicit titles for each dataset within the \change{Multiples} Panel, to help orient \change{analysts} in their search for attributes. These Concerns were specified through a third manually specified configuration file, with one line for each Concern listing the attributes that should appear in each; attributes could appear in more than one group.

\subsubsection{S4 Customize}
\label{sec:architecturalEvolution:controllabilityEvolution:s4Customize}
The fourth stage brought interactive control to data customization (\task{Customize}), allowing \change{analysts} to explicitly manage during runtime what had previously been manually specified by the developer and loaded at startup time. While inspecting a single attribute (\task{Inspect}) in the Detail Panel, \change{analysts} could filter out (or keep only) specific levels of that attribute, merge levels together, and rename them. These changes were made persistent with the first use of non-specified configuration files: the interaction logs generated in one session were automatically saved and re-applied at startup time in subsequent sessions. 

Further customization (\task{Customize}) support was also added through specified configuration, to support color assignments for specific attribute levels and ordering assignments across levels within specific attributes. We had noticed several instances of Stroop effect  \cite{Macleod1991} confusion when demonstrating previous prototypes to stakeholders: triage categories had colors implicit in their names, like \texttt{HLBCYellow}, so it was disorienting for them to be rendered in any other color. Similarly, they understandably expected ordered attributes such as survey responses to be shown in numerical order. We chose configuration as a fast way to address these particular requirements in a more general and flexible way than simply hardwiring these through code. 

Control of partitioning (\task{Partition}) was augmented through a combination of interaction and configuration. The previous mechanism of implicitly inferring whether to inspect, partition, or stratify based on the type and number of the selected attributes was conceptually parsimonious and thus quick to program, but was difficult to understand and confusing for \change{analysts}. We changed the interface so that users made this choice explicitly through radio buttons to provide feedback on the current action mode and affordances for how to change it. It also allowed users to interactively choose the granularity with which to partition the data over time. By adding specified configuration of which attribute within each of the four datasets corresponded to time, the interface could support partitioning by requiring \change{analysts} to interactively select only a single attribute, for reduced cognitive load. 

\subsubsection{S5 Curate}
\label{sec:architecturalEvolution:controllabilityEvolution:s5Curate}
The fifth stage augmented configurable curation of attribute groups (\task{Curate}) with interactive control, adding interface support for \change{analysts} to remove or add attributes to Concerns and move them around within it, and also make copies of and rename Concerns. Although our original vision for Concerns was that \change{analysts} would select the relevant one for their tasks, we found they typically used the default one containing all attributes and very rarely changed to a different pre-defined Concern. However, they more actively engaged with this mechanism when provided with interactive controls for doing so. Moreover, we repurposed this mechanism to support new functionality both interactively and through logs. The Explode action, introduced in this stage as an interactive choice, automatically creates a new Concern populated with the newly generated attributes in addition to the source attributes. This interaction logs are also saved and re-applied as a configuration file on subsequent startups.

\subsubsection{S6 Final}
\label{sec:architecturalEvolution:controllabilityEvolution:s6Final}
The visual interface evolved noticeably in this final step, with the addition of the new chart type of Sankey diagrams to show flow and many low-level usability improvements of the interface with widget layout and labelling. However, there was no change of controllability methods during this final stage of \toolname{}'s evolution. Interactive control had been achieved for all six tasks at the previous stage. We did not identify any compelling reasons to add further configuration, either by specifying or logging. 

\subsection{Software Architecture}
\label{sec:architecturalEvolution:softwareArchitecture}

The architecture of \toolname{} follows the standard division into front-end client and back-end server. The implementation uses Python supported by Flask for the back-end server and JavaScript supported by D3.js for the front-end client. All configuration files are stored on disk and loaded into memory by the back end, and are passed to the front end when requested. Many files consist of a first section of manual specifications generated by the developer that is included with the software distribution and saved at install time, followed by a second section where logged information implicitly or explicitly specified by \change{users'} interaction logs is appended with each subsequent usage session. 

One constraint with many architectural implications was data security requirements. The data was available on a file server under the administrative control of the stakeholders and shared with the visualization team. We chose to avoid the substantial engineering effort of a permanently running server with stringent authentication of user credentials, allowing us to stay completely out of the data stewardship loop. 

Although it was permissible for the stakeholders themselves to use data internally in many ways, the data sharing agreement dictated that no healthcare data should ever be saved to local disk by the visualization researchers. \toolname{} uses a local server, typically running only for the duration of the client session. On startup, a simple data upload page on the front end allows users to specify the location of the shared data files that were uploaded to the back end and parsed on the fly (shown in Supplemental Figure S12). A straightforward way to support customization and curation was to have \toolname{} operate purely locally, so that each user's configuration files were simply stored with their existing personal computer account. The configuration files contain only metadata, no patient data. 

The need to customize data (\task {Customize}), starting at stage S4, led to the challenge of how and where to store data modifications such as merging attribute levels together. Our solution, which dovetailed nicely with our approach to controllability through configuration, was to treat all data modifications made through customization as a series of operations applied sequentially to the base version of the datasets. We could thus store the operation logs on local disk to preserve that information across sessions, separately from the base data itself, which is not saved. We held two full copies of the data in memory in the back-end server: one untouched and one with the operations applied to it. 

This architecture also enabled straightforward undo operations in the back end and front end. In the back end, we could simply clone the untouched version of the data and apply the modified set of operations to it. In the front end, \toolname{} includes a separate window to explicitly inspect the sequence of data customization operations and undo any of them on demand, shown in Supplemental Figure S13. The Operation History Panel provides a complete provenance trail of all customizations applied to the data as a simple table in reverse-chronological order, in the spirit of log file analysis. 
The last operation can be undone by clicking the undo button at the top. The user can also select any operation from the list for deletion, highlighting it and any later operations that depend on it. We did not pursue the full level of polish for undoing operations as we did in the main interactive interface as only highly technical end users used this page and we did not observe sufficient interest in it from the stakeholders.

\section{Related Work}
\label{sec:relatedWork}

We discuss the related work in the healthcare domain, data wrangling, visualization tools and techniques, and visualization methodology. 

\subsection{Healthcare}
\label{sec:relatedWork:healthCare}

\toolname{} has a natural connection to the healthcare domain, but differs considerably from prominent types of healthcare visualization approaches.
In line with important surveys, we structure the discussion by a) the clinical treatment of individual patients~\cite{shneidermanPH13}, b) the tracking of diseases~\cite{preimL20}, and c) the analysis of patient cohorts~\cite{rindWAMWPS13,westBH15}.

Approaches for the \emph{treatment of individual patients} include the collection, observation, and analysis of personal health information~\cite{senorAT12,faisalBP13}. 
Typical user groups include patients collecting sensor data with small mobile devices for self-analysis~\cite{patientsLikeMe,quantifiedSelf} and physicians making clinical decisions about patient treatment
~\cite{plaisantMRWS96,mamykinaGHB04,mueller2020}.
In contrast to approaches for individual patients, \toolname{} provides decision-making support at the health system level of resource usage by large volumes of patients.

Applications for the \emph{tracking of diseases} emphasize single diseases by taking information of multiple patients and other sources into account. We differentiate between public health 
and epidemiological approaches. A common theme in public health approaches is using interactive dashboards to track and model individual diseases or problems, such as Covid-19, or dashboards for injuries or infectious disease models~\cite{comba20}, often following the goal to prevent diseases by informing the general public and political decision makers~\cite{afzalME11,burmeister2021}. In contrast, the focus of epidemiological approaches is on creating evidence in determinants of health and disease conditions~\cite{preimKHHOTV16,alemzadehHNCIVS17}, often facilitated with confirmatory data analysis and hypothesis validation.
In contrast, \toolname{} tracks overall health system resources, rather than a particular disease. Moreover, the geographic emphasis of disease tracking is a mismatch for our target \change{analysts}. 

The \emph{analysis of patient cohorts} is a third stream of healthcare approaches, which can further be subdivided into patient stratification, exploratory patient analysis, and relation seeking.
Patient stratification approaches often provide an overview of electronic health records of patients and allow the grouping of patients into cohorts by meaningful attributes~\cite{cga2015}. Exploratory patient analysis systems help to reveal structures, patterns, and outliers in electronic health records~\cite{angelelliOHTHLLPH14,tvcg2018Medi}, or to compare cohorts~\cite{malikDMOPS15,glueckGCKBW17}. 
Focusing on longitudinal aspects in electronic health records~\cite{wongsuphasawat2011outflow,gotz2014decisionflow,caballero2017visual} and identifying local temporal patterns~\cite{wongsuphasawatGPWTS11,monroe2013temporal,guoJGDZC19} are frequently applied approaches that also fall into that category. 
Applications for relation-seeking tasks support the identification of relations, correlations, dependencies, and associations between attributes in patient data, e.g., for the analysis of health and population surveys~\cite{torresEBEM12,klemmLGNHVP16}, diseases like asthma or cancer~\cite{bhavnani2014role,cgf2014jb}, or medical imaging data~\cite{bannach2017}.
\toolname{} differs from these patient analysis approaches in the goal to preserve health care system capacity for delivering care, rather than exploring (temporal) disease patterns.
\toolname{} further supports monitoring resource usage to assess impacts and detect unanticipated patterns with an emphasis on patient attachment and satisfaction, rather than direct patient treatment.

\subsection{Data Wrangling}
\label{sec:relatedWork:dataWrangling}
\toolname{} enables experts to conduct data wrangling~\cite{Kandel2011} operations to cope with data quality challenges and make data more useful for individual analysis goals.
A long list of general purpose tools for data wrangling exists, including OpenRefine~\footnote{OpenRefine, https://openrefine.org}, Trifacta Wrangler~\cite{KandelPHH2011}, or Alteryx~\footnote{Alteryx, https://www.alteryx.com}. 
These tools are useful if data wrangling is executed as a dedicated pre-process.
However, the \toolname{} use case requires data wrangling support as an interactive and integral step in the analysis workflow, with a seamless back-and-forth between data wrangling and data analysis~\cite{Kandel2011,KandelPHH2012}, which makes general purpose tools inappropriate. 
In the healthcare domain, inspiring approaches exist that combine data wrangling and downstream data analysis, but differ in their application focus, such as the analysis of patient cohorts with electronic health records~\cite{burmeister2019self} or the analysis of cohort study data in epidemiology~\cite{AlemzadehNIVSSP17}.
A second difference of \toolname{} is the specific need of wrangling capability for only a small subset of the general case so that using general purpose tools may be overkill.
In specific, \toolname{} supports a) wrangling of attribute categories (levels) including filtering (in/out), merging, renaming, as well as partition/stratify/bin operations, and b) wrangling of attributes such as rename and duplicate.
In contrast, there is no need to support direct wrangling of attribute values. 


\subsection{Visualization Tools and Techniques}
\label{sec:relatedWork:visualizationTools}
Several types of tools allow the creation of interactive dashboards~\cite{SarikayaCBTF2019}, either with a business analytics tools focus~\cite{behrischSSSMWMP19} such as Tableau, Looker, or PowerBI, with grammars such as Vega~\cite{vega} and Vega-lite~\cite{vega-lite}, or with task-based dashboard creation engines like QualDash~\cite{elshehalyRBMAGR21}.
These tools offer fully customizable encodings of data attributes and the fully flexible creation of dashboards, with a major learning curve. 
\looseness = -1
\toolname{} differs from these approaches in three ways. First \toolname{} is designed for a user group that needs only a very small set of fixed encodings, with no need for customization, and no need for chart composition. Second, \toolname{} needs to support only carefully curated small set of tasks (show attribute change over time, stratify one attribute by another). Finally, \toolname{} supports the need of experts for interactive data wrangling throughout the interactive analysis workflow, not just as pre-processing step.

The design of the \toolname{} visual encoding is further inspired by various visualization techniques. One need was the visualization of flows to observe changes in triage assessments between nurses and physicians, so we use Sankey-style diagrams for triage flow views~\cite{wongsuphasawat2011outflow}, in a simple setting of only two or three flow stages.
\looseness = -1
The coordination of views in \toolname{} is inspired by van den Elzen's and van Wijk's small multiples and large singles approach~\cite{elzenW13}. Their approach shows small multiples with alternatives from a current state (large single). Our adaptation is slightly different: we show a large single view, as the result of an attribute selection from small multiples, showing either a subset of all attributes or an attribute grouping. 


\subsection{Visualization Methodology}
\label{sec:relatedWork:designStudyMethodology}

Our proposed Controllability Through Configuration \change{(CTC)} model is a methodological contribution. 
Many design and process-oriented methods have been proposed for visualization design, including design study methodology~\cite{SedlmairMM12}, data-driven design studies~\cite{OppermannM2020}, design study lite methodology~\cite{SyedaMRBB2020}, and design activity framework~\cite{McKennaMAM2014}. In contrast, \change{the CTC model} that we propose focuses on architectural considerations. Similarly, the HCI literature on rapid ethnography~\cite{Millen2020} 
and rapid contextual design~\cite{HoltzblattWW2005} \change{is} also more focused on design methods  than software architecture considerations. 
\change{The CTC model is not a direct alternative to these design models, but rather a cross-cutting proposal: it provides scaffolding and guidance on architectural choices that would allow developers to carry out any of them.}
\change{Techniques for how to carry out specific stages within our  CTC model} are discussed at length in the literature, e.g., logging~\cite{HeerMSA2008} and interacting~\cite{YiKSJ2007}. In contrast, our contribution is focused on how to move between these stages systematically. 

\section{Discussion}
\label{sec:Discussion}

Here, we reflect on the design study, paying particular attention to the challenges we faced.

\subsection{Case Studies As Validation Evidence}

The intention of the first case study, the chauffeured demo run by DigEM's KH, was to demonstrate the utility of \toolname{} to HLBC stakeholders. To achieve this goal, \change{the tool builders on the design team} took part in crafting the narrative of the demo to ensure that it contained all of the key features of \toolname{}. \change{Although that narrative is the synthesis of many previous analyses using evolving versions of the prototype, rather than representing a single usage session, it is structured around questions posed by a domain expert (KH) with deep knowledge of the entire HEiDi project.} 

The second case study was not a demo, but a test drive by HN, a front-line analyst at DigEM. Although this test drive was guided by real analysis questions, HN had been significantly involved in the design of \toolname{} and had attended many chauffeured demo sessions prior to the test drive. Her involvement in the project means that her chosen analysis questions and her use of the tool would have likely been influenced by the ways she had previously seen the design team use it.

These first two case studies provided evidence that our tool, \toolname{}, was successfully able to answer the analysis questions that we had designed it for in collaboration with DigEM domain experts. However, \change{they do not validate whether these questions were} representative of the real analysis that was conducted by HLBC analysts. 

The third case study helps to resolve this issue. EC, an HLBC analyst, had not used \toolname{} prior to the session, and had only heard of it from colleagues. Despite this lack of experience with the tool, his use of it was quite similar to KH's use in the chauffeured demo and HN's use in the test drive. This match reveals that our design supports the real analysis tasks of the HLBC analysts and that that the first two case studies, although influenced by the design team, were accurate demonstrations of conducting analysis tasks with \toolname{}. \change{Moreover, the finding that \toolname{} allowed for analysis in substantially less time than the previous workflow was important confirmation of utility.}

\change{Our choice of validation through qualitative case studies arose from two principles. First, the general methodological considerations of design studies suggest that case studies of real users, real problems, and real data are by far the most common way to validate this kind of work~\cite{SedlmairMM12}. Second, the particular context of a staged design process imposes even more stringent constraints on what kind of validation is feasible, because time with the intended target users is a very scarce resource~\cite{McLachlanMKN2008}. Securing the time of HLBC analyst EC to use \toolname\ for the final case study was only possible after successfully demonstrating \toolname{}'s value to HLBC leadership, via the first case study. We deemed a more formal study focused on quantitative metrics to be a poor fit with this situation.}


\subsection{\change{Next Steps Not Taken}}

\change{After Case Study 3, the HLBC leadership expressed significant interest in the software, inquiring about both expansion to larger datasets and longer-term deployment within their organization. We interpret this enthusiasm for \toolname{} as further indications that our design was successful in addressing target user needs and that targeted analysis tools could have a place in the organization.}
\change{However, we realized that to support usage by multiple front-line analysts across the full volume of HLBC calls, while continuing adherence to the privacy protocols for public health data, was not feasible with our existing research prototype: it would have required multiple person-years of work from a development team.} 

\change{As mentioned in Section~\ref{sec:processAndAbstractions:dataContext}, \toolname{} only includes data from encounters that are sent to physicians from nurses, only 7\% of total HEiDi calls; moreover, that total is only 1\% of the total volume of HLBC calls.} We considered that level of enthusiasm to be a heartening validation of utility; \change{such a deployment would have allowed us to conduct longitudinal studies. However,} scaling up \toolname{} to handle a scope \change{multiple orders of magnitude beyond the original target} would \change{have required not only} significant software engineering for both scale and robustness, but a full architectural redesign. The full scale of their data is not simply a question of volume, but of heterogeneity: they collect many datasets of different formats and semantics in addition to the four data sources that are currently incorporated. Moreover, our research prototype does not have the level of robustness required for a production codebase; the engineering effort to create a system sufficiently robust that it could be deployed to a large audience in the healthcare sector would be considerable. 
\change{Our decision was that we had already learned enough about the design of targeted analysis tools that further polishing of the prototype would not have sufficient research benefits to be worth the time costs. We thus turned our efforts to fleshing out the Controllability through Configuration model.}
 
\change{In choosing to end the project at this point, we grappled with what Meyer and Dykes call the \textit{ethics of exit}~\cite{meyer2019criteria} and what Akbaba \etal~call \textit{troubling matters of care}~\cite{akbaba2023troubling} within design study collaborations. Their discussion of \textit{neglected things} speaks to this mismatch between the resources required to conduct a paper-sized visualization research project, vs.~to provide a data analysis tool robust enough for long-term use in a public health setting.}  

\subsection{Color Challenges}
\label{sec:Discussion:colorChallenges}

HLBC's internal use of color names for triage dispositions and our use of color in \toolname{}'s visualization design led to the possibility of the Stroop Effect \cite{Macleod1991} occurring, where a mark's encoded color does not match the color in the name of the associated item. We noticed the Stroop Effect early on and were able to easily ensure that attribute levels in the original datasets were given matching colors through configuration. Once new attributes were derived with new levels, however, the configuration of the original levels was no longer applied, leading to the mismatches shown in Figures~\ref{fig:uc3T5MD}a and b. 

We considered more complex back-end systems, such as one that would scan level names for color words; however, this approach would not have  sufficed as some levels would have associated colors that would not be named, such as the level of \texttt{MD-TryHomeTreatment} in Figure~\ref{fig:UC3DispositionOverTime} which is given the color green. Our configuration strategy resolved almost all Stroop Effect situations. The more bespoke solution of making color modifiable through user interaction would have required substantial development time,  required some learning time from \change{analysts}, and would have affected only a handful of situations, so we did not choose to support that level of interactivity in \toolname{}.

\subsection{Data Challenges}
\label{sec:Discussion:dataChallenges}

We handled several challenges regarding data, some anticipated and some unexpected.

We followed two of the recommendations provided by Crisan et al \cite{CrisanGM2016}, who discuss the significant regulatory and organizational constraints in domains such as healthcare, where sensitive data entails 
significant restrictions on who can access it and where it can be stored. First, we successfully built trust through a staged design process, where delayed access to data was planned for and obtained -- reasonably quickly by the standards of the field, within four months. These datasets were very straightforward to de-anonymize with minimal processing, in contrast to other more complex cases. Second, and most crucially, we successfully identified the impact of constraints on data access very early on, namely that data should never be saved locally by visualization researchers. This realization led us to develop a somewhat unusual yet lightweight and flexible architecture that guaranteed this data security policy by design. Otherwise, we might have taken a more conventional approach, and then needed to backtrack to achieve that requirement. 


The third challenge involved changes to both the data itself and its format during the project. As expected, the size of the datasets steadily increased, with collection of data after each day's service. Although all dataset sizes were still well within the limits of manageability, the streaming nature of the data precluded any kind of overall computation in advance, in contrast to a static situation.
A more difficult issue is that the format of the data would occasionally change, in response to both the evolution of the project itself and changes in the collection process. Most commonly, these structural changes were due to added, removed, changed, or re-ordered attributes. The flexibility of our design approach of using configuration files to handle data processing meant that we were able to quickly update our system to work with the new data format, despite not \change{initially knowing} the data format would change.

\subsection{Assumptions versus Reality}
\label{sec:Discussion:assumptions}
We reflect on \change{two initial assumptions that} did not hold.

One initial assumption was that deriving new attributes should be considered a first-class citizen, so an initial version of our task abstraction included it as an explicit task to support.
However, observing analysts use the tool revealed that, while deriving new attributes occurs in the tasks \task{Customize} (in entirety) and \task{Stratify} (in parts), it was not the primary goal of the analysts; it was a means, rather than an end \cite{Brehmer2013}. 
Rather than fully deriving new secondary attributes, we identified a stronger focus on processing attributes. Thus, we do not include Derive as an explicit task in our final task abstraction. Instead, we felt that the \change{Adapt} framing of Customize and Curate was a better match for the mental models of the analysts. 

Our second assumption was that analysts would invest time and effort in an initial cleaning stage, and then benefit from that effort being amortized over many subsequent analysis phases. Many visualization tools are built to support this kind of two-phase wrangling and pre-processing model, where early-stage tasks to check and ensure the validity of the data are followed by later-stage tasks pertaining to deeper analysis questions \cite{Meyer2009}. However, in this project, the reality of usage patterns appeared to be different: experts did not choose to preserve cleaned data states for later re-use, but instead re-executed data cleaning operations anew during each analysis session. In \toolname{}, while re-using previously cleaned data is an available feature, it is not required; none of our users chose to do so. We have several conjectures as to why this assumption did not hold. It is certainly possible that they were simply unaware that \toolname{}~ functionality included the ability to re-use previously cleaned data; we did not explicitly suggest that users do so.   
One intriguing possibility is that they perceived making permanent changes to data as unwise. Crisan et al \cite{CrisanGM2016} note the prevalence of in healthcare settings of institutional policies that correcting errors and other forms of data cleaning are only carried out by selected authorized individuals, to ensure a systematic and consistent process. It could be that these norms are so ingrained for these analysts that re-starting from scratch in each analysis session was the most natural way to proceed. It is also possible that they were reluctant to change current data 
in fear
that the streaming nature of the data would lead to inconsistencies with data that arrived later; although our architecture was intended to handle this case cleanly, by applying configuration rules at startup time, we did not emphasize this capability to the analysts. 
Another possibility is that they did not think that investing time into data cleaning would pay off, in the absence of a definite commitment to using this tool for ongoing analysis needs in the future. 

\subsection{\change{Generalizability and Transferability}}

\change{\toolname{} itself has only been demonstrated in a single specific healthcare setting, but we conjecture that it could be used in a broad variety of other contexts both within and beyond healthcare. Although the goals are very specific to this setting, our task abstractions and the interactive workflow may be transferrable. The six abstract tasks (\emph{Inspect}, \emph{Partition}, \emph{Stratify}, \emph{Scan}, \emph{Customize} and \emph{Curate}) are completely domain-agnostic. The underlying assumptions of the interactive \textbf{Scan, Act, Adapt} workflow instantiated in \toolname{} are less general yet still domain-agnostic: they should transfer to any other domains where individual analysis questions require considering a few attributes at once, situated within a larger analysis scope with many dozens of already-familiar attributes. We would like to investigate the transferability to other domains in future work.} 

The Controllability Through Configuration model has only been applied to a single case of a single domain so far, so validating the extent to which this idea generalizes will depend on the results of future use. We note that this proposed approach to conducting design studies has no domain-specific aspects at all, so we believe it could be beneficial in future visualization work across a wide range of application domains. We are enthusiastic about testing the generalizability and transferability of this configuration-based approach in other contexts. 
We consider it a useful mechanism for instantiating \textit{data sketches}, as proposed by Lloyd and Dykes \cite{LloydD2011}, who suggest that showing users their data in the system must be done as early as possible; we advocate that the Controllability Through Configuration model allows for high flexibility with low development time.

\subsection{\change{Why Not Existing Applications}}

\change{Custom research prototypes are costly to develop, and as discussed above have limits to their robustness. An obvious question is whether existing off-the-shelf applications such as Tableau or PowerBI, which have had immense engineering resources dedicated to their reliability and scalability, would have been sufficient to solve this problem. One pragmatic obstacle that we heard about from the HLBC analyst (EC) is that these existing business intelligence tools did not comply with their specific data privacy requirements. A deeper consideration is to consider this question through the lens of the \textbf{relevance ratio}: what fraction of the capabilities of a general-purpose tool are relevant vs.~irrelevant for the use case that we identified. In this case, the vast majority of the capabilities of these existing systems are not relevant.} 

\change{As we discuss in Sec~\ref{sec:processAndAbstractions:designConsiderations}, our design target is the middle of the \emph{Flexibility} axis, in between the general-purpose tools like Tableau and a highly specialized tool. \toolname{} and its  \textit{Scan, Act, Adapt} workflow abstraction aim at a small set of very abstract tasks, serving as a technology probe for an interesting region of the analysis space differs from the capabilities supported by standard tooling.} 



\subsection{Limitations and Future Work}
\label{sec:Discussion:futureWork}

We chose to limit the interactivity of \toolname{} to only what was most necessary, by beginning with low interactivity and iteratively adding more in the areas that mattered most to \change{analysts}. Just as with our decision to not implement interactive color modification (Section~\ref{sec:Discussion:colorChallenges}), there are many other obvious possibilities where richer interaction could be beneficial, even though we did not deem them important enough to warrant significant development time during this design study. One is the ability to interactively re-order levels within an attribute, for example to change the ordering of the binned client ages of \texttt{Young}, \texttt{Old}, \texttt{Middle} in Figure~\ref{fig:teaser}b to the more natural order of \texttt{Young}, \texttt{Middle}, \texttt{Old}, is one of them. Another is to support more complex attribute derivations by adding the ability to apply formulas and perform calculations on attributes.
However, although it would certainly be possible to build a tool with much richer functionality, we note the benefits of controlling the scope of the project. We made a quite deliberate choice to halt the iterative refinement process at the current point, which we consider a sweet spot in the tradeoff space of functionality vs.~development effort. In this sense, the limitations and the strengths of the project are two sides of the same coin.



\section{Conclusion}

We \change{have presented} data and task\change{, and workflow} abstractions for the exploration and analysis of HLBC's virtual healthcare service delivered by nurses and physicians, abstracted from multiple meetings with domain experts. We then designed and developed \toolname{}, a visual analytics tool based on these abstractions, to support the improvement of this service. To validate \toolname{}, we conducted three case studies with DigEM and HLBC staff. Through our design process, we developed the Controllability Through Configuration model to support rapid iteration of visualization software, which is particularly appropriate in design studies. We encourage further use of this model as a \change{generalizable} approach to visualization design iteration. 

\ifCLASSOPTIONcompsoc
  \section*{Acknowledgments}
\else
  \section*{Acknowledgment}
\fi

The authors wish to thank HLBC and the BC Ministry of Health for the conceptual support of this study and the permission to use anonymized data securely \change{for this} study. This work was supported in part by a grant from the UBC Data Science Institute. We appreciate helpful comments from Steve Kasica, Zipeng Liu, Michael Oppermann, and Ben Shneiderman. 

\ifCLASSOPTIONcaptionsoff
  \newpage
\fi



\bibliographystyle{IEEEtran}
\bibliography{IEEEabrv, template}
%

%







\def\interBioSpace{-33pt} 
\vspace{\interBioSpace}
\begin{IEEEbiography}[{\includegraphics[width=1in,height=1.25in,clip,keepaspectratio]{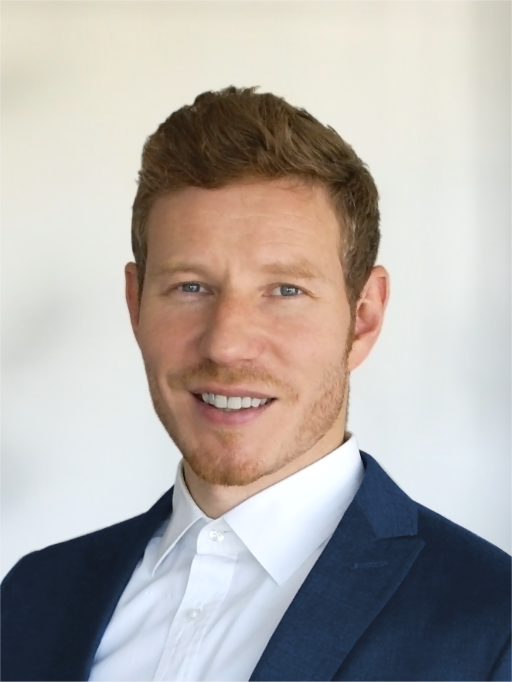}}]{J{\"u}rgen Bernard}
is an assistant professor at the University of Zurich and head of the Interactive Visual Data Analysis group. He has received the PhD degree in computer science from the University of Darmstadt, and was a postdoctoral research fellow at the University of British Columbia, Canada. His research focus is on visual analytics and human-centered machine learning.
\end{IEEEbiography}
\vspace{\interBioSpace}

\begin{IEEEbiography}[{\includegraphics[width=1in,height=1.25in,clip,keepaspectratio]{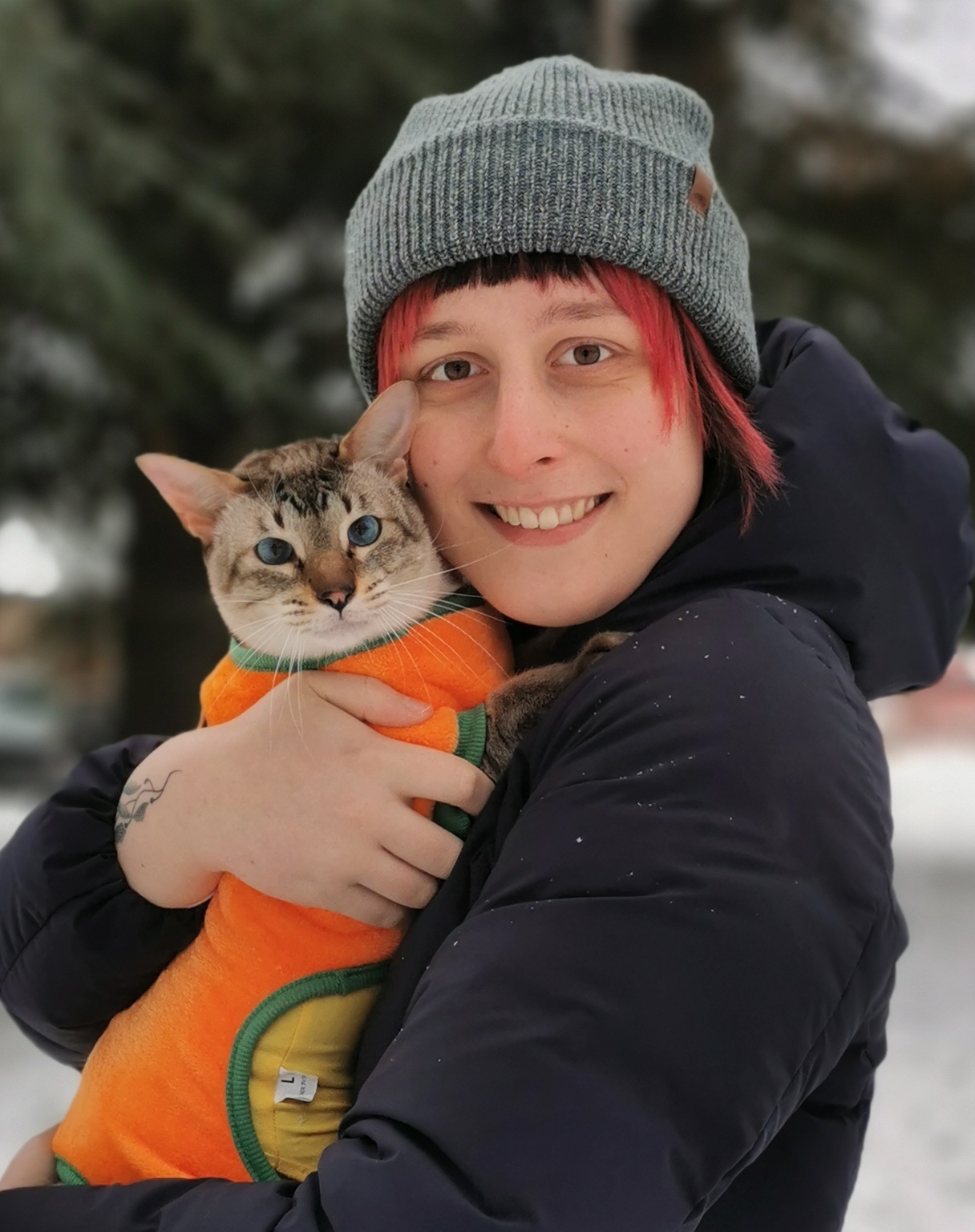}}]{Mara Solen}
is a PhD candidate at the University of British Columbia, supervised by Tamara Munzner. Her current research focus is on visualization for everyday people, delivered through avenues such as education and communication.
\end{IEEEbiography}
\vspace{\interBioSpace}

\begin{IEEEbiography}[{\includegraphics[width=1in,height=1.25in,clip,keepaspectratio]{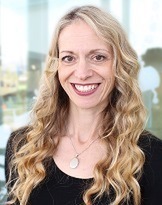}}]{Helen Novak Lauscher}
is Associate Lead at UBC Digital Emergency Medicine, where she collaborates with a team working on research and evaluation in digital health, virtual care, and patient and public engagement in health contexts from acute to community. With a background in educational and counselling psychology, she specializes in participatory approaches to research and evaluation, qualitative methods, and using digital health tools to build capacity.
\end{IEEEbiography}
\vspace{\interBioSpace}

\begin{IEEEbiography}[{\includegraphics[width=1in,height=1.25in,clip,keepaspectratio]{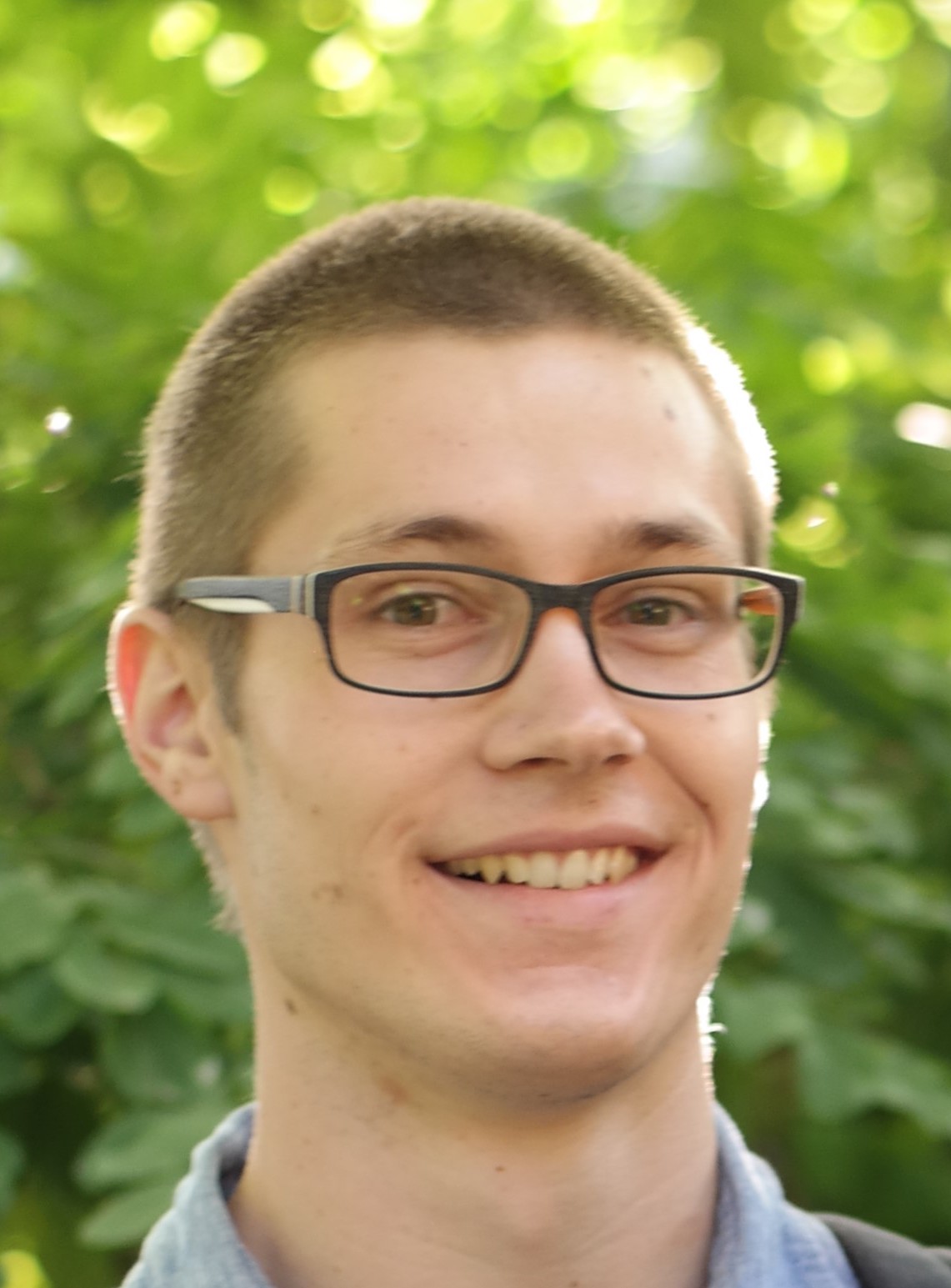}}]{Kurtis Stewart}
is a senior research analyst at the UBC Digital Emergency Unit. His background is in psychology and research methodology in interdisciplinary studies. His current work involves data wrangling, analysis, and visualization for provincial virtual care programs and health system evaluation.
\end{IEEEbiography}
\vspace{\interBioSpace}

\begin{IEEEbiography}[{\includegraphics[width=1in,height=1.25in,clip,keepaspectratio]{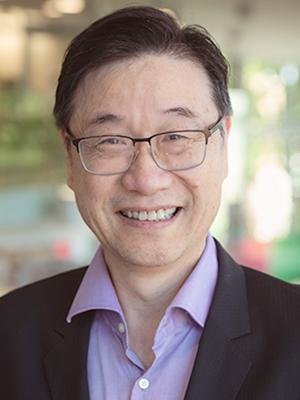}}]{Kendall Ho}
(MD, FRCPC, FCAHS, MGC)
is an emergency physician specialist, a professor at the Department of Emergency Medicine, University of British Columbia, and medical director of the British Columbia Ministry of Health HealthlinkBC Virtual Physician Services. He leads the UBC Digital Emergency Medicine Unit, which explores the use of digital technologies and data science to enhance health services and support patients in emergency departments, and their transition between home and hospital care.
\end{IEEEbiography}
\vspace{\interBioSpace}

\begin{IEEEbiography}[{\includegraphics[width=1in,height=1.25in,clip,keepaspectratio]{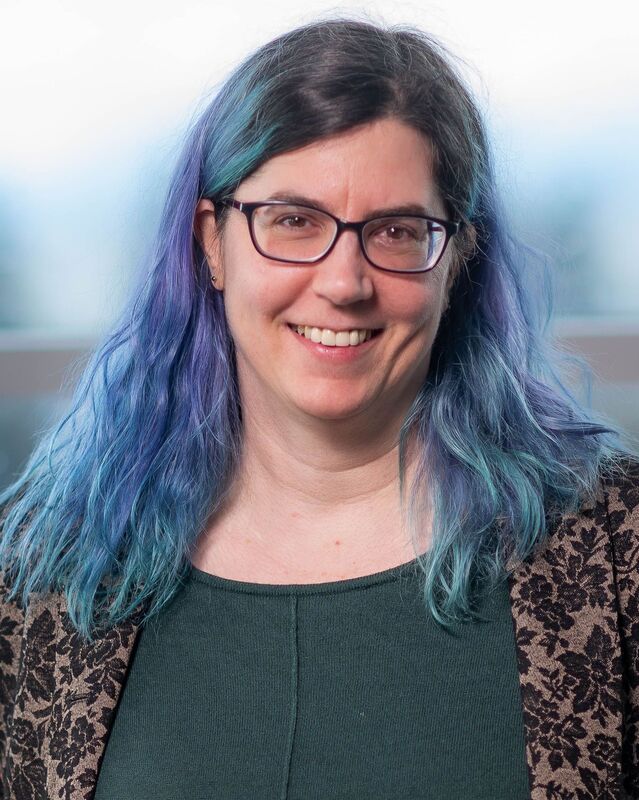}}]{Tamara Munzner}
(IEEE Fellow) received the PhD degree from Stanford. She is currently a professor with the University of British Columbia. She has worked on visualization projects in a broad range of application domains from genomics to journalism. Her book Visualization Analysis and Design is heavily used worldwide, and she was the recipient of the IEEE VGTC Visualization Technical Achievement Award.
\end{IEEEbiography}
\vfill

\end{document}